\documentclass[10pt,twocolumn]{article}

\usepackage[english]{babel}

\usepackage[letterpaper,top=2cm,bottom=2cm,left=2cm,right=2cm,marginparwidth=1.75cm]{geometry}

\usepackage[colorlinks=true, allcolors=blue]{hyperref}
\pdfoutput=1
\usepackage[]{natbib}
\usepackage[nottoc]{tocbibind}
\usepackage{minted}
\usepackage{algorithm}
\usepackage[noend]{algpseudocode}
\usepackage{braket}
\usepackage[utf8]{inputenc}
\usepackage[T1]{fontenc}
\usepackage{hyperref}
\usepackage{graphicx}
\usepackage[babel]{microtype} 
\usepackage[bb=boondox]{mathalfa}
\usepackage{amsmath,amssymb,amsthm,bm,amsfonts,mathrsfs,bbm}
\usepackage{caption}
\usepackage{subcaption}
\usepackage{abstract}
\usepackage{makecell}

\usepackage{authblk}

\title{\large \textbf{Hardware-inspired Continuous Variables Quantum Optical Neural Networks}}

\author[1,2]{Todor Krasimirov-Ivanov\protect\footnotemark[1]}
\author[1]{Alba Cervera-Lierta}
\author[1]{Paolo Stornati}
\author[3,4]{Federico Centrone}

\affil[1]{Barcelona Supercomputing Center (BSC), 08034 Barcelona, Spain}
\affil[2]{Universitat de Barcelona, 08007 Barcelona, Spain}
\affil[3]{ICFO--Institut de Ciencies Fotoniques, The Barcelona Institute of Science and Technology}
\affil[4]{Universidad de Buenos Aires, Instituto de Física de Buenos Aires (IFIBA)}
\date{}

\begin{document}
\begingroup
\renewcommand\thefootnote{\fnsymbol{footnote}}
\twocolumn[
    \vspace*{-2em}
    \maketitle
    \vspace{-3em}
    \begin{onecolabstract}
        Continuous-variables (CV) quantum optics is a natural formalism for neural networks (NNs) due to its ability to reproduce the information processing of such trainable interconnected systems. In quantum optics, Gaussian operators induce affine mappings on the quadratures of optical modes while non-Gaussian resources---the challenging piece for physical implementation---originate the nonlinear effects, unlocking quantum analogs of an artificial neuron. This work presents a novel experimentally-feasible framework for continuous-variable quantum optical neural networks (QONNs) developed with available photonic components: coherent states as input encoding, a general Gaussian transformation followed by multi-mode photon subtractions as the processing layer, and homodyne detection as outputs readout. The closed-form expressions of such architecture are derived demonstrating the family of adaptive activations and the quantum-optical neurons that emerge from the amount of photon-subtracted modes, proving that the proposed design satisfies the Universal Approximation Theorem within a single layer. To classically simulate the QONN training, the high-performance \href{https://github.com/bsc-quantic/QuaNNto}{\texttt{QuaNNTO}} library has been developed based on Wick--Isserlis expansion and Bogoliubov transformations, allowing multi-layer exact expectation values of non-Gaussian states without truncating the infinite-dimensional Hilbert space. Experiments on supervised learning and state-preparation tasks show balanced-resource efficiency with strong expressivity and generalization capabilities, illustrating the potential of the architecture for scalable photonic quantum machine learning and for quantum applications such as complex non-Gaussian gate synthesis.
    \end{onecolabstract}
    \vspace{1em}
    ]
    \footnotetext[1]{todor.krasimirov@bsc.es}
\endgroup

\section{Introduction}

Information processing in artificial neural networks (NNs) is based on adjusting mathematical---linear and nonlinear---transformations stacked in a layered pattern that act on input data entries, aiming to obtain the best approximation to the corresponding predictions of such inputs in the output readout of the network. This training process generates a NN model which, with some accuracy, generalizes the response domain of a target problem by capturing patterns from some of its samples collected in the training dataset.

Continuous-variable (CV) quantum optics serves as a natural environment to implement analogs of classical neural networks by exploiting the infinite-dimensional Hilbert space of optical modes. In the scheme of CV quantum optical neural networks (QONN), Gaussian unitaries realized with displacements, squeezers, beam splitters and phase shifters lay the role of affine transformations on the optical quadratures---continuous degrees of freedom of photons---, while non-Gaussian resources supply the essential activation---nonlinear---function and universality for CV quantum information processing \cite{lloyd1999quantum}.

Early CV-QONN proposals realized \emph{neurons} by alternating Gaussian layers with explicit non-Gaussian unitaries as \emph{Kerr} gate, thereby enforcing nonlinearity and universality of CV quantum computing \cite{lloydcvqnn,steinbrecher2019quantum}. A complementary direction pursued measurement-induced nonlinearity: Siopsis \emph{et al.} used only Gaussian unitaries with ancillas and a homodyne repeat-until-success protocol to impart non-Gaussian kicks to the data modes \cite{siopsis2023exprealcvqnn}, while Chabaud \emph{et al.} studied adaptive linear optics where each layer applies an interferometer followed by photon counting with feed-forward \cite{chabaud2021quantum}.

More recently, programmable nonlinear meshes have been introduced by embedding tunable Kerr-like elements directly into interferometer networks \cite{chernykh2024quantum}, and particle-number preserving photonic convolutional NNs have employed state injection of non-Gaussian ancilla at designated layers \cite{monbroussou2025photonic}. 

\begin{figure*}[ht!]
    \centering
    \includegraphics[width=2\columnwidth, angle=0]{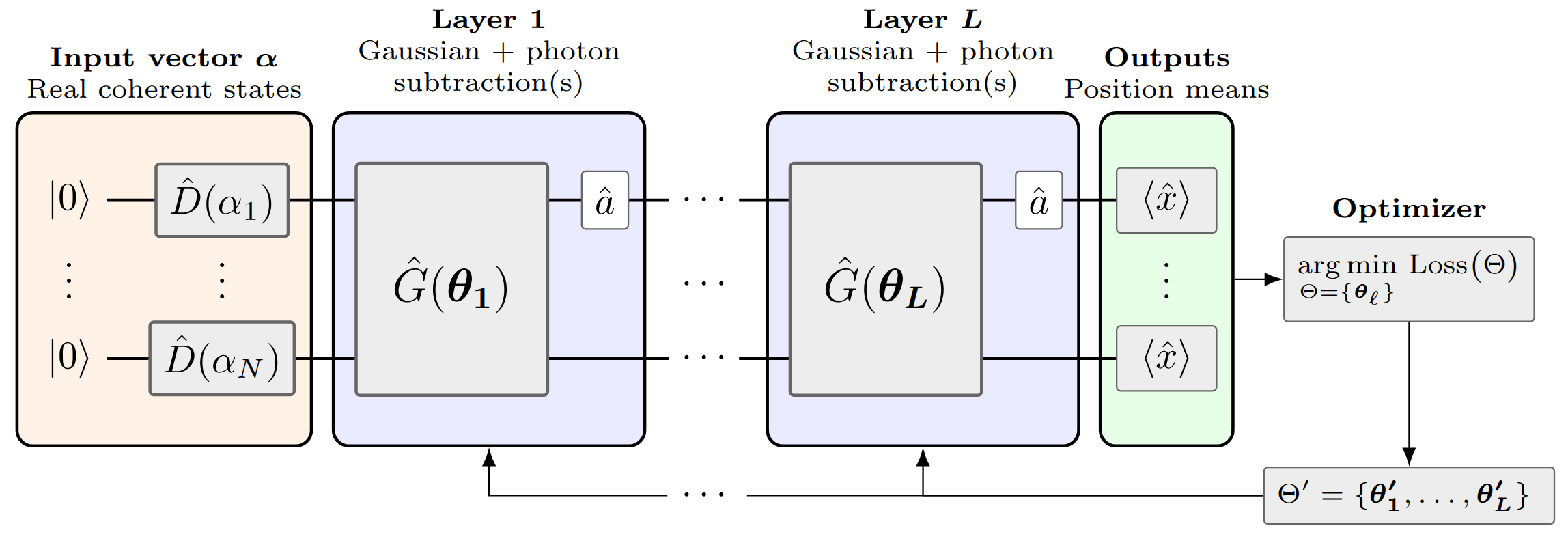}
    \includegraphics[width=2\columnwidth, angle=0]{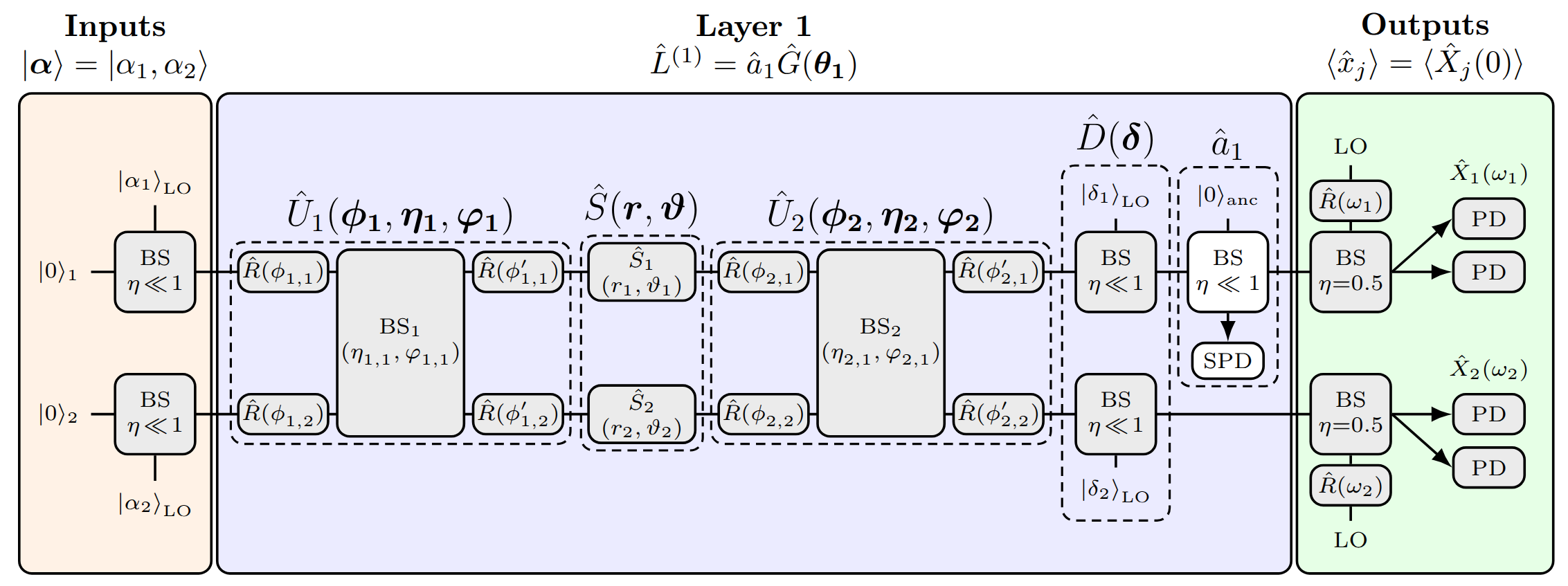}
    \caption{General CV-QONN architecture with single photon subtraction per layer. Top: $N$-mode QONN with $L$ layers where each layer is carried out by the operator $\hat L = \hat a_1 \hat G = \hat a_1 \hat{D} \hat{U}_2 \hat{S} \hat{U}_1$. Bottom: Hardware implementation example of a single-layer two-mode QONN using local oscillators (LO), beam splitters (BS), phase shifters (PS or $\hat R$), single-mode squeezers ($\hat S$), single-photon detectors (SPD) and photodetectors (PD).}
    \label{fig:physical-qnn}
\end{figure*}

The practical development of a CV-QONN hinges on balancing experimental overhead with the non-Gaussian resources needed for universality. Non-Gaussian unitaries such as the Kerr gate, while universal in principle, are widely regarded as hardware-demanding at optical frequencies because the third-order susceptibility \(\chi^{(3)}\) yields very small per-photon phase shifts, making high-fidelity deterministic implementations challenging \cite{shapiro2006single}. Measurement-induced cubic operations are generally possible, but they require tailored non-Gaussian ancilla states and high-performance feed-forward, which remains experimentally demanding \cite{marek2011deterministic}. However, the state-engineering literature established \emph{heralded photon subtraction} as a practical non-Gaussian resource: tapping a weak fraction of squeezed light and conditioning on a photon detection event reliably produces non-Gaussian states such as cat states and odd-photon superpositions \cite{ourjoumtsev2006generating,neergaard2006generation,wakui2007photon,olivares2005squeezed}. Recent approaches, such as generalized photon subtraction with photon-number-resolving (PNR) detection, further improve cat-state generation rates and operating regimes \cite{takase2021generation}.

Inspired by the variational CV circuit of Stornati \emph{et al.} \cite{stornati2024variational}, this work presents the CV-QONN architecture from Fig. \ref{fig:physical-qnn} where, in a quantum system of several optical modes, initial coherent states encode the input, parameterized Gaussian optics---interferometers $\hat U(\bm\phi,\bm\eta,\bm\varphi)$, squeezers $\hat S(\bm r,\bm\vartheta)$, and displacements $\hat D(\bm\delta)$---are assembled in a general Gaussian operator $\hat G(\bm\theta)$ and realize the affine part of each layer, photon subtraction $\hat a$ provides the nonlinearity, and homodyne detection of position quadrature $\langle \hat x\rangle$ implements the output readout. The entire stack of quantum operators maps directly onto hardware-friendly components: local oscillators, beam splitters, phase shifters, optical parametric amplifiers for single-mode squeezers, single-photon detectors and photodetectors.

A key feature of the introduced design is that, for quantum states built with Gaussian operators and a finite number of annihilation/creation operators, expectation values of observables admit exact analytic expressions via Bogoliubov transformations and Wick--Isserlis expansion in terms of the first and second statistical moments of the underlying Gaussian state \cite{weedbrook2012gaussian, ulysse2021classical}. This allows closed-form expression derivations on genuinely non-Gaussian states generated by photon subtraction/addition operations, which enable gradient-based training and avoids Fock-space cutoffs that are common in classical simulations of CV quantum systems.

\section{Overview}
The proposed CV-QONN design from Fig. \ref{fig:physical-qnn} demonstrates a hardware-feasible framework for implementing NN-like computation using available photonic components. Each QONN layer is constructed from a tunable full-system Gaussian unitary followed by multi-mode photon subtractions. This layer processing together with input encoding in initial coherent states and output obtaining via homodyne detection yield analytic, closed-form expressions for the subtraction nonlinearity, the overall layer mapping, and the complete operation of a multi-layer QONN through Bogoliubov transformations and Wick--Isserlis expansions.

The main theoretical result is the derivation of the closed-form expression in Eq. \eqref{ph-sub-expval}---generalized for multi-mode subtractions on a general Gaussian state in Eq. \eqref{qonn-layer-output}---that establishes the emergence of non-polynomial, tunable ridge activations induced by photon subtraction on squeezed coherent states. This analytic mapping reveals that each photon-subtracted mode corresponds to an effective quantum-optical neuron. Consequently, a single QONN layer with linear readout satisfies the Universal Approximation Theorem (UAT) on compact sets \cite{cybenko1989approximation, leshno1993multilayer}. When several subtractions occur over correlated squeezed modes, additional nonlinear cross-terms appear, enhancing the expressivity of the network.

From a practical viewpoint, all required photonic components---beam splitters, phase shifters, squeezers, heralded photon-subtraction modules, and homodyne detection---are experimentally accessible with current technology. Limitations such as the available squeezing strength ($|r| \leq 1.7$) \cite{vahlbruch2024squeezinglims} are mitigated by the tunable affine mapping stage, ensuring the inputs lie within the effective nonlinear regime.

To classically simulate and train these QONNs, the high-performance \texttt{QuaNNTO} library was developed, leveraging the Wick--Isserlis expansion and Bogoliubov transformations to build analytic expressions of QONN setups with various neurons and layers. Such expressions allow to compute exact expectation values of non-Gaussian states without truncation in the infinite Hilbert-space of the quantum system, thus, enabling gradient-based training. Simulations were performed on the MareNostrum 5 supercomputer at the Barcelona Supercomputing Center.

Several benchmark experiments confirm the model’s expressivity and generalization capabilities:
\begin{itemize}
    \item \textbf{Curve fitting:} single-layer QONNs with photon subtraction accurately reproduce complex nonlinear functions (e.g., polynomial, exponential, oscillatory) from noisy datasets, whereas purely Gaussian networks yield only linear fits.
    \item \textbf{Classification:} nonlinear decision boundaries such as in the ``moons'' and ``circles'' datasets are correctly learned by single-neuron QONNs with two and three modes, achieving perfect accuracy. A multi-class MNIST classification task with continuous encodings confirms scalability to realistic data.
    \item \textbf{Non-Gaussian gate synthesis:} the QONN successfully learns to reproduce the action of complex non-Gaussian operators such as the cubic phase gate $V(\gamma) = e^{i \gamma x^3/3}$ by matching high-order statistical moments of the gate effect on coherent states.
\end{itemize}

These results collectively show that the QONN architecture merges analytic tractability, experimental feasibility, and high representational power. Its universal single-layer design, together with exact classical simulation and gradient-based training, positions it as a promising candidate for scalable photonic quantum machine learning and quantum information processing tasks such as quantum state engineering and non-Gaussian gate synthesis.

Additionally, the provided analytic formalism is shown to be applicable beyond photon subtraction, admitting generalization for different architectures whose operators structure can be implemented via ladder polynomials and Gaussian operators, such as weak Kerr interactions.

\section{Background}
This section gathers the foundations of classical NNs and CV quantum optics needed to build a bridge between the operational block of feed-forward NNs and CV quantum optical states and operators.

\subsection{Feed-forward neural networks}
The mathematical formalism of classical feedforward NNs is the main reference for designing an effective CV-QONN architecture.

Let $\bm \alpha \in \mathbb{R}^{A}$ denote the input vector and $\Phi:\mathbb{R}\!\to\!\mathbb{R}$ be a scalar activation function. For each output $j \in \{1,...,M\}$, a single-layer of $K$ hidden units in a fully-connected NN implements the operation 
\begin{equation}
\label{nn_singlelayer}
\mathcal L_j(\bm \alpha) \;=\; \bm c_j+\sum_{k=1}^{K} V_{jk}\,
\underbrace{\Phi \big(W_k\bm \alpha +\bm d_k\big)}_{\text{ridge function}},
\end{equation}
where $W \in \mathbb{R}^{K\times A}$ is the weight matrix that linearly transforms $\bm{\alpha}$ along with the bias vector $\bm d \in \mathbb{R}^K$, while $V \in \mathbb{R}^{M \times K}$ and $\bm c \in \mathbb{R}^M$ provide the linear readout of the layer, yielding the outputs $\{\mathcal L_j(\bm \alpha)\}_{j=1}^M$. Each term $\Phi(W_k \bm \alpha + \bm d_k)$ depends on the input only through the one-dimensional projection $W_k\bm \alpha$, which consequently corresponds to a ridge function. The amount of ridge functions $K$ in the layer determines the number of neurons and therefore the width of the layer. This enables the Universal Approximation Theorem (UAT) when $\Phi$ is non-polynomial, meaning that a single-layer NN can approximate any continuous function with a decreasing error when growing the width of the layer \cite{cybenko1989approximation, leshno1993multilayer}.

Deep NNs are built stacking $L$ of such layers whose output $\tilde f^{(L)}(\bm{\alpha})$ is equal to the composition 
\begin{equation}
\label{nn_multilayer}
    \tilde f^{(L)}(\bm{\alpha}) = \mathcal{L}^{(L)} \circ \cdots \circ \mathcal{L}^{(1)}(\bm{\alpha}).
\end{equation}

Training a multi-layer NN involves tuning the parameters $\{W^{(\ell)},\bm d^{(\ell)},V^{(\ell)},\bm c^{(\ell)}\}_{\ell=1}^{L}$ in order to approximate its output $\tilde f^{(L)}(\bm{\alpha})$ to a given target function $f(\bm{\alpha})$ representing the problem to be learned. The activation function $\Phi$ is typically fixed and allows the NN to learn patterns beyond linearity.

\subsection{CV quantum optics}
A CV quantum optical system composed by $N$ modes can be described either in phase-space via the quadratures or canonical operators $(\hat x, \hat p)$ of each optical mode, or in Fock space via their field or ladder operators $(\hat a, \hat a^\dagger)$ \cite{weedbrook2012gaussian}
\begin{equation}
\begin{split}
\label{s_vec}
    &\bm{\hat s} = (\hat{x}_1, \hat{x}_2, ..., \hat{x}_N, \hat{p}_1, \hat{p}_2, ..., \hat{p}_N)^{\mathsf T} \\
    &\bm{\hat b} = \bar{U} \hat{\bm s} = (\hat{a}_1, \hat{a}_2, ..., \hat{a}_N, \hat{a}^\dagger_1, \hat{a}^\dagger_2, ..., \hat{a}^\dagger_N)^{\mathsf T},
\end{split}
\end{equation}
where 
\begin{equation}
    \bar{U} = \frac{1}{\sqrt{2}}
    \begin{pmatrix} \mathbb{1}_N & i\mathbb{1}_N \\ \mathbb{1}_N & -i\mathbb{1}_N\end{pmatrix},
\end{equation}
is the change of basis matrix that brings the system from phase-space representation to Fock space.

For $\hat{\bm s}$ or $\hat{\bm b}$ to represent physical quantum systems, the equivalences 
\begin{equation}
\begin{split}
\label{ccr}
    &[\bm{\hat{s}}, \bm{\hat{s}}^\dagger] =
    iJ\text{ with }J = \begin{pmatrix}
        \mathbb{0}_N & \mathbb{1}_N \\ -\mathbb{1}_N & \mathbb{0}_N
    \end{pmatrix},\\
    &[\bm{\hat{b}}, \bm{\hat{b}}^\dagger] = i\bar{U}J\bar{U}^\dagger = \begin{pmatrix} \mathbb{1}_N & \mathbb{0}_N \\ \mathbb{0}_N & \mathbb{-1}_N\end{pmatrix}
\end{split}
\end{equation}
must hold to preserve the canonical commutation relations (CCR) in the system \cite{serafini2017quantum}. For the purpose of this work, the characterization of quantum optical systems will be carried out in Fock space with $\bm{\hat b}$.

The states of CV quantum systems are divided into two regimes: Gaussian and non-Gaussian quantum states.

On the one hand, Gaussian quantum states are referred to systems governed by, at most, quadratic Hamiltonians and whose phase-space of their quadratures can be represented by a Gaussian distribution \cite{serafini2017quantum}.

An $N$-mode quantum system in any pure Gaussian state $\rho_G$ is fully and uniquely characterized by its first two statistical moments: the means vector $\bm{\bar b}$ and the covariance matrix $\Sigma$, defined as
\begin{equation}
\begin{split}
    \rho_G \equiv
    \begin{cases}
        \bm{\bar{b}} = \langle \bm{\hat{b}} \rangle \\
        \Sigma = \langle \bm{\hat{b}} \bm{\hat{b}}^\dagger \rangle - \langle \bm{\hat{b}} \rangle \langle \bm{\hat{b}}^\dagger \rangle
    \end{cases}
\end{split}
\end{equation}
implying that classical systems can efficiently simulate Gaussian quantum states \cite{weedbrook2012gaussian, serafini2017quantum}.

Gaussian operators correspond to unitary operators in Fock space and affine transformations in phase-space. Any Gaussian transformation acting on a Gaussian state preserves the system's Gaussianity, which also implies that the product of Gaussian operators yields another Gaussian operator \cite{weedbrook2012gaussian, ferraro2005gaussian}. 

For a system of $N$ optical modes, Gaussian operators generated by linear Hamiltonians correspond to an assembly of $N$ single-mode displacement operators $\hat{D}_{\bm \delta} = \hat{D}(\bm \delta) = \bigotimes_{k=1}^N\hat{D}_k(\delta_k),\ \bm \delta=(\delta_1,\dots,\delta_N) \in \mathbb{C}^N$ whose action over $\rho_G$ statistics is
\begin{equation}
\label{disp_op}
    \hat{D}_{\bm \delta} \rho_G \hat{D}_{\bm \delta}^\dagger:
    \begin{cases}
        \bm{\bar{b}} \rightarrow \bm{\bar{b}} + \bm d \\
        \Sigma \rightarrow \Sigma
    \end{cases}\text{with } \bm d = (\bm\delta, \bm\delta^\ast)^{\mathsf T}.
\end{equation}

Conversely, any Gaussian operator $\hat Q$ generated by a purely quadratic Hamiltonian is completely defined by its Bloch-Messiah decomposition $\hat{Q} = \hat{U}_2 \hat{S} \hat{U}_1$, where $\hat{U}_1$ and $\hat{U}_2$ are $N$-mode passive interferometers---set of beam splitters (BS) and phase shifters (PS)---, and $\hat{S}$ is an assembly of $N$ single-mode squeezers. $\hat Q$ acts on $\rho_G$ statistics, preserving the CCR of Eq. \eqref{ccr}, through its associated Bogoliubov matrix $B$ as
\begin{equation}
\begin{split}
\label{quad-gaussian-op}
    &\hat{Q} \rho_G \hat{Q}^\dagger:
    \begin{cases}
        \bm{\bar{b}} \rightarrow B\bm{\bar{b}} \\
        \Sigma \rightarrow B\Sigma B^\dagger
    \end{cases} \text{where}\\
    &B = \begin{pmatrix}
        U & V \\
        V^\ast & U^\ast
    \end{pmatrix} \text{ and } \begin{cases}
        U = U_2  \cosh R\ U_1^\dagger\\
        V = -U_2 e^{i2\varTheta} \sinh R\ U_1^{\mathsf T}
    \end{cases},
\end{split}
\end{equation}
with $U_1,U_2 \in U(N)$ representing the interferometers, and $R=\text{diag}(r_1,\dots, r_N) \geq 0$ and $\varTheta = \text{diag}(\vartheta_1,\dots,\vartheta_N)$ describe optical modes squeezing factors and phases respectively.

Thus, the most general $N$-mode Gaussian operator $\hat G$ is defined by joining both linear and quadratic Gaussians \cite{weedbrook2012gaussian} in the sequence
\begin{equation}
\label{general-gauss-lin-quad}
    \hat G = \hat{D}_{\bm \delta} \hat{Q} = \hat{D}_{\bm \delta} \hat{U}_2 \hat{S} \hat{U}_1.
\end{equation}

On the other hand, non-Gaussian quantum states cannot be identified only by their first two statistical moments, and, in the general case, their classical tractability goes beyond efficient Gaussian simulation. They are a necessary resource to achieve universal quantum computation in CV systems \cite{walschaers2021non}.

Non-Gaussian operators, under some conditions, tend to introduce statistical correlations of order higher than two and can inject nonlinear phase-space dynamics into the quantum state they act on \cite{lloydcvqnn}, analogous to activation functions in the classical NN framework.

In contrast to physically demanding non-Gaussian operators such as Kerr or cubic-phase gates, this work uses a hardware-accessible method to generate non-Gaussianity in a quantum optical system via the non-unitary particle-annihilation---photon subtraction---operator $\hat a$ \cite{biagi2025photon}.

For a non-Gaussian state built as a polynomial of ladder operators acting on some pure Gaussian state $\rho_G$, Wick--Isserlis theorem decomposes any expectation value of such state construction into a sum of products of linear and quadratic ladder expectations, which can be collected from the means and covariances of $\rho_G$ \cite{stornati2024variational, walschaers2023emergent}. The general Wick--Isserlis expansion for non-zero means \cite{nosal2022higher} applied to this framework is defined as
\begin{equation}
\begin{split}
\label{perf_match}
    &Tr[\hat{a}^\#_{m_1}...\hat{a}^\#_{m_M} \rho_G] = \\
    &\sum_{\mathcal{P}_j \in \mathcal{P}} \prod_{ \{\hat p_1,\hat p_2\} \in \mathcal{P}_j }
    \begin{cases}
        Tr[\hat p_1\rho_G]=f(\bm{\bar{b}}) \text{ if } \hat p_1 = \hat p_2 \\
        Tr[\hat p_1 \hat p_2\rho_G]=f(\Sigma) \text{ if } \hat p_1 \neq \hat p_2 
    \end{cases}
\end{split}
\end{equation}
where $\hat a^\#$ can be either the creation or annihilation operator, $m_j$ is the mode the ladder operator acts on, and $\mathcal{P}$ is the set of sets containing all existing perfect matchings with self-loops combinations from the set of ladder operators $\{\hat{a}^\#_{m_1},...,\hat{a}^\#_{m_M}\}$.

For an expression with $M$ ladder operators, the total amount of different matchings is given by
\begin{equation}
\label{num_lpms}
    |\mathcal{P}| = \sum_{k=0}^{\lfloor M/2 \rfloor} \frac{M!}{2^k \cdot k! \cdot (M - 2k)!}.
\end{equation}

\section{CV-QONN architecture}
This section formally justifies the effectiveness and power of the proposed QONN architecture shown in Fig. \ref{fig:physical-qnn} that mimics the operation of a NN layer from Eq. \eqref{nn_singlelayer} through a strategic assembly of the aforementioned quantum optical operators. The closed-form characterizations of the transformations performed by the QONN layer are derived along with the input uploading and the output acquisition schemes.

\subsection{Gaussianity as affine mapping}
\label{sec:linearity}
Let $\bm \alpha \in \mathbb{R}^M$ be the $M$-dimensional input vector to be uploaded, $\hat G$ from Eq. \eqref{general-gauss-lin-quad} be the $N$-mode general Gaussian unitary, and Eqs. \eqref{disp_op} and \eqref{quad-gaussian-op} be the actions of Gaussian transformations over an arbitrary $N$-mode Gaussian state. The $N$-mode quantum system starts in the vacuum---Gaussian---state $\rho_\text{v}$ defined by the null vector of means $\bm{\bar{b}}_\text{v}$ and the normalized identity covariance matrix $\Sigma_\text{v}$ \cite{weedbrook2012gaussian} like
\begin{equation}
\label{vacuum-st}
    \rho_\text{v}=\ket{0}\bra{0}^{\otimes N} \equiv
    \begin{cases}
        \bm{\bar{b}}_\text{v} = \bm 0^{2N}\\
        \Sigma_\text{v} = \frac{1}{4} \mathbb{1}_{2N}
    \end{cases}.
\end{equation}
Then, loading the inputs $\bm \alpha$ on $\rho_\text{v}$ as real coherent states
\begin{equation}
\label{input-loading}
    \rho_{\bm \alpha} = \hat{D}_{\bm{\alpha}} \rho_\text{v} \hat{D}_{\bm{\alpha}}^\dagger \equiv
    \begin{cases}
        \bm{\bar{b}}_{\bm \alpha} = (\bm \alpha, \bm 0^{N-M},\bm \alpha^\ast,\bm 0^{N-M}) \\
        \Sigma_{\bm \alpha} = \Sigma_\text{v}
    \end{cases},
\end{equation}
and applying $\hat G$, the final Gaussian state becomes
\begin{equation}
\label{affine-stats}
    \rho_G = \hat G \rho_{\bm \alpha} \hat G^\dagger \equiv
    \begin{cases}
        \bm{\bar b} = B\bar{\bm b}_{\bm \alpha} + \bm d\\
        \Sigma = B \Sigma_0 B^\dagger.
    \end{cases}
\end{equation}
The means $\bm{\bar b}$ of the optical modes in $\rho_G$ carry the affine mapping of the inputs $B\bm \alpha+\bm d$ and can be trained by adjusting the free parameters of $\hat G$ as seen in Fig. \ref{fig:physical-qnn}, which are unfolded in terms of physical components as $\bm\theta = \{\bm\phi_1, \bm\eta_1, \bm r, \bm\vartheta, \bm\phi_2, \bm\eta_2, \bm\delta\}$ for each QONN layer.

\subsection{Subtraction as nonlinear mapping}
\label{sec:nonlinearity}
In this section, the characterization of the nonlinearities that emerge from the photon subtraction operator is analytically derived, proving precisely the achievement of an effective and tunable activation function. Also, this behavior is generalized to multi-mode quantum systems with diverse subtractions, coupling it to the concept of quantum neuron in the QONN.

\begin{figure*}[ht!]
    \centering
    \includegraphics[width=\columnwidth, angle=0]{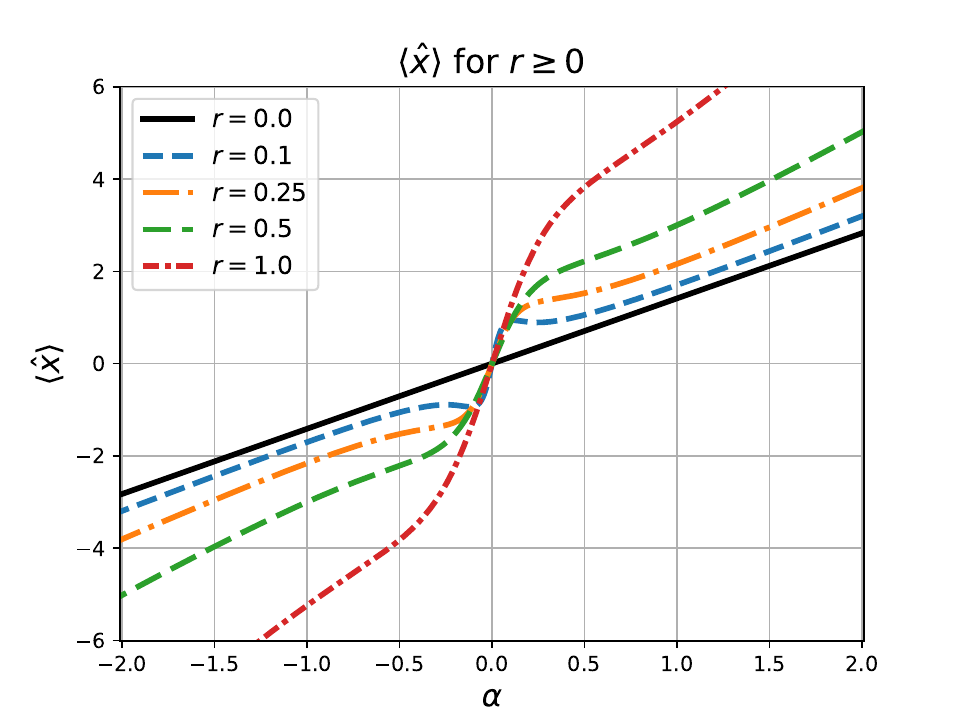}
    \includegraphics[width=\columnwidth, angle=0]{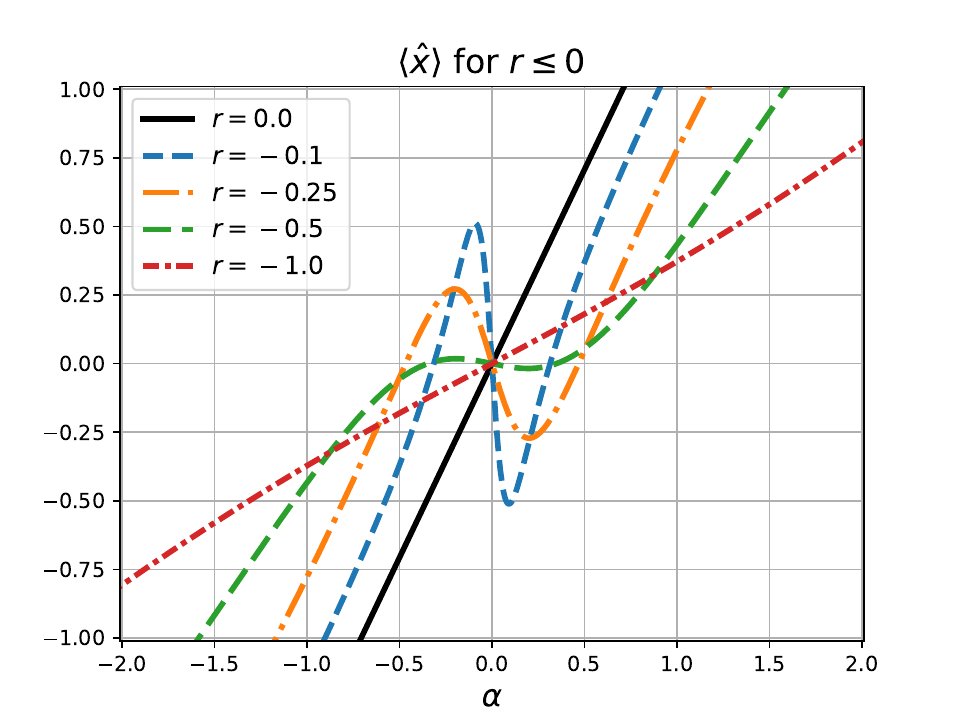}
    \caption{Nonlinearity of the photon subtraction in the QONN: expectation values of Eq. \eqref{ph-sub-expval} with respect to real displacements $\alpha \in (-2, 2)$ for different factors of (left) momentum squeezing and (right) position squeezing.}
    \label{fig:ph-subtract-sq-coherent-states}
\end{figure*}

So far, the input is encoded in the QONN as a coherent state $\rho_{\bm \alpha}$ described by Eq. \eqref{input-loading}, the operator $\hat G$ from Eq. \eqref{general-gauss-lin-quad}---responsible for the affine mapping---contains squeezing as the active optical component, and the outputs are obtained by quadrature means through homodyne measurements. Thus, analyzing the effect of the annihilation operator on the quadratures of squeezed-coherent states is key to formalizing its induced nonlinearity and seeing the activation form. Consider the single-mode non-Gaussian state 
\begin{equation}
\label{ph-sub-coherent-wf}
\ket{\psi(\alpha,r)} = \hat a\, \hat S(r) \,\hat D(\alpha)\ket{0} = \hat a \ket{r\alpha} \text{ with } \alpha, r \in \mathbb{R},
\end{equation}
where a sequence of displacement, squeezing ($\vartheta = 0$), and annihilation operators act on the vacuum state. The final Gaussian state $\rho_{r\alpha} = \ket{r\alpha} \bra{r\alpha}$ is
\begin{equation}
    \rho_{r \alpha} \equiv 
    \begin{cases}
        \bm{\bar{b}} = (e^r \alpha, e^r \alpha^\ast) \\
        \Sigma = \begin{pmatrix}
            \cosh^2r & \frac{1}{2} \sinh 2r\\
            \frac{1}{2} \sinh 2r & \sinh^2 r
        \end{pmatrix}
    \end{cases},
\end{equation}
so, expanding by Wick, the normalized expectation value of the mode's position quadrature is worked out as
\begin{equation}
\begin{gathered}
\label{ph-sub-expval}
\Phi_r(\alpha) := \langle \hat x \rangle_{\psi ({\alpha,r})} = 
\sqrt{2} \Re \Big(\frac{\text{Tr}[\hat a^\dagger \hat a \hat a \rho_{r\alpha}]}{\text{Tr}[\hat a^\dagger \hat a \rho_{r\alpha}]}\Big) = \\
\sqrt{2} \Re \Big(\frac{\langle\hat a^\dagger \hat a\rangle\langle \hat a \rangle + \langle\hat a^\dagger \hat a\rangle\langle \hat a \rangle + \langle\hat a \hat a\rangle\langle \hat a^\dagger \rangle + \langle\hat a^\dagger \rangle\langle \hat a\rangle\langle \hat a \rangle}{\langle \hat a^\dagger \hat a \rangle + \langle \hat a^\dagger \rangle \langle \hat a \rangle}\Big)=
\\
\sqrt{2} \Re \Big(\bm{\bar{b}}_1 + \frac{\Sigma_{22}\bm{\bar{b}}_1 + \Sigma_{12}\bm{\bar{b}}_2}{\Sigma_{22} + |\bm{\bar{b}}_1|^2 }\Big)=\\
\sqrt{2}\Big(\underbrace{e^r \alpha}_{\text{Linear map }L_r(\alpha)} + \underbrace{\frac{\alpha \cdot e^{2r} \sinh r}{\alpha^2 \cdot e^{2r} + \sinh^2 r}}_{\text{Nonlinear bump }R_r(\alpha)}\Big)
\end{gathered}
\end{equation}
where $\langle \cdot \rangle = \bra{r\alpha}\cdot\ket{r\alpha}$ are expectation values over $\rho_{r\alpha}$ omitted for brevity, and $\Re$ denotes the real part. Fig. \ref{fig:ph-subtract-sq-coherent-states} illustrates some shapes chosen from the continuous set of activations that this expression offers through the parameter $r$ with respect to the real input $\alpha$.

The following properties of Eq. \eqref{ph-sub-expval} are reasonably found:
\begin{enumerate}
    \item \textit{Odd.} $\Phi_r(-\alpha) = - \Phi_r(\alpha)\ \forall \alpha,r$
    \item \textit{Smooth.} For $r \neq 0$, the denominator of $R_r(\alpha)$ satisfies $\alpha^2 \cdot e^{2r} + \sinh^2 r > 0\ \forall \alpha$, and  $\Phi_{r=0}(\alpha) = \Phi_{r\rightarrow0^-}(\alpha) = \Phi_{r\rightarrow0^+}(\alpha)= \sqrt{2} \alpha$, implying smoothness and continuity.
    \item \textit{Non-polynomial.} For $r\neq 0$, $R_r(\alpha)$ is a rational bump $\forall\alpha$, making squeezing an essential ingredient for nonlinearity injection.
    \item \textit{Asymptotic linearity and nonlinear saturation.}\\
    For $|\alpha| \gg 1$:
    \begin{equation}
    \label{linear-alpha-inf}
        \Phi_r(\alpha) = \sqrt{2} e^r \alpha + O\Big(\frac{1}{\alpha}\Big),
    \end{equation}so the nonlinear term $R_r(\alpha)$ vanishes for large $|\alpha|$ and the map tends to the linear asymptote $L_r(\alpha)$.
    \item \textit{Nonlinear range for $\alpha$.} The vanishing of $R_r(\alpha)$ in Eq. \eqref{linear-alpha-inf} for large $|\alpha|$ implies that such a displacement needs to be restricted to a range where the photon subtraction has a non-negligible activation effect on $\alpha$ when retrieving the output. Then, to maintain the ratio between $L_r(\alpha)$ and $R_r(\alpha)$ above a given threshold $\varepsilon$ like
    \begin{equation}
    \tau_r(\alpha) := \Big|\frac{R_r(\alpha)}{L_r(\alpha)}\Big| = \frac{e^{r}|\sinh r|}{\alpha^{2}e^{2r}+\sinh^{2}r} \geq \varepsilon,
    \end{equation}
    $\alpha$ must be bounded within 
    \begin{equation}
    \label{alpha-nonlinear-range}
        |\alpha| \leq e^{-r}\,
     \sqrt{\frac{e^{r}|\sinh r|}{\varepsilon}-\sinh^{2}r}=\alpha_{\max}(r,\varepsilon).
    \end{equation}
    Therefore, for any fixed $r \neq 0$ and $0<\varepsilon<\tau_r(0)$, the set of $\alpha$ such that $\tau_r(\alpha) \geq \varepsilon$ is the finite interval $|\alpha|\leq \alpha_{\max}(r,\varepsilon)$. Outside this range, the nonlinear correction of the subtraction is smaller than a fraction $\varepsilon$ of the linear term.
    \item \textit{Squeezing effect.}\\
    For $r=0$, $\Phi_0(\alpha) =\sqrt{2}\alpha$ is linear.\\
    For $r \rightarrow \infty$, $\sinh r \sim e^r/2$ and the relative nonlinear contribution is
    \begin{equation}
    \label{sq-inf}
         \tau_r(\alpha) = \frac{2}{1+4\alpha^2} + O(e^{-2r}),
    \end{equation}\\
    which is bounded between $0$ and $2$ and depends only on $\alpha$, not on $r$. Hence, increasing the squeezing parameter $r$ beyond a moderate value does not enhance the degree of nonlinearity of the map: the bump saturates and only its overall amplitude is rescaled by the exponential factor $e^r$.\\
    For $r \rightarrow -\infty$, $\sinh r \sim -e^{-r}/2$, then
    \begin{equation}
    \label{sq-minusinf}
        \tau_r(\alpha) = \frac{2e^{2r}}{1+4\alpha^2 e^{4r}} + O(e^{4r}) \sim 2 e ^{2r} \rightarrow 0,
    \end{equation}
    so the nonlinear contribution disappears exponentially for large negative squeezing.\\ Consequently, a moderate squeezing factor $r \neq 0$ is sufficient to effectively induce nonlinearity in position quadrature through photon subtraction.
    \item \textit{Position vs. momentum squeezing.} For real input $\alpha$ and squeezing angle $\vartheta = 0$, the sign of relative nonlinear correction $\frac{R_r(\alpha)}{L_r(\alpha)}$ is determined by $\sinh r$, so it flips when squeezing position ($r<0$, $\hat x$ squeezed) instead of momentum ($r>0$, $\hat x$ anti-squeezed). The magnitude of the nonlinearity $\tau_r(\alpha)$ in the map $\alpha\mapsto\langle\hat x\rangle$ becomes strongly curved (sigmoidal-like) for moderate and large positive $r$, whereas for large negative $r$ the nonlinear bump is suppressed and the response tends to a linear function of $\alpha$. Thus, photon subtraction induces qualitatively different nonlinear activations depending on whether position or momentum is squeezed.
\end{enumerate}

These features make Eq. \eqref{ph-sub-expval} a suitable activation function for the QONN whose bump gain and form can be controlled by $r$, providing a complete family of nonlinear mappings of the input $\alpha$ in the $\hat x$ quadrature as exemplified in Fig. \ref{fig:ph-subtract-sq-coherent-states}, and enabling the UAT \cite{cybenko1989approximation, leshno1993multilayer}.

Additionally, appendix \ref{app:photon-addition-nonlinearity} provides the same activation function analysis using photon additions instead, which also gives place to a suitable nonlinearity that satisfies the UAT. However, the tunability in the additions case is limited with respect to subtractions when it comes to squeezed coherent states, where the set of nonlinear mappings is richer.

\subsection{CV-QONN layer closed form}
The assembly of the QONN layer consists of joining the linear mapping seen in section \ref{sec:linearity} together with the nonlinearity proven in section \ref{sec:nonlinearity}. Then, the closed-form expression of the layer is provided by deriving the expectation values representing the QONN outputs. To do so, the nonlinear mapping of Eq. \eqref{ph-sub-expval} has to be extended from the photon subtraction effect on a single-mode squeezed coherent state to subtractions over a multi-mode general Gaussian state.
Let 
\begin{equation}
\label{multi-subtraction-op}
    \displaystyle \hat A_K = \left(\prod_{k \in K}\hat{a}_k\right)
\end{equation}
be a sequence of photon subtraction applied over different optical modes contained in the set $K$.
Then, the proposed architecture for the CV-QONN layer is defined as the quantum operator 
\begin{equation}
\label{gen_qnn_layer}
   \hat L = \hat A_K \hat G = \hat A_K \hat{D}_{\bm \delta} \hat{U}_2 \hat{S} \hat{U}_1.
\end{equation}

The final normalized quantum state of a single-layer QONN is defined as 
\begin{equation}
\label{final-state-qonn}
    \rho = \frac{\hat L \rho_{\bm \alpha} \hat L^\dagger}{\text{Tr}(\hat L \rho_{\bm \alpha} \hat L^\dagger)} = \frac{\hat A_K \rho_G \hat A_K^\dagger}{\text{Tr}\Big[\hat A_K^\dagger \hat A_K \rho_G \Big]},
\end{equation}
which, by homodyne detection in each mode $j$, yields the QONN outputs
\begin{equation}
\label{expval-qonn}
    \langle \hat x_j \rangle_\rho = \sqrt 2 \Re\ \text{Tr}[\hat a_j \rho] =  \sqrt 2 \Re \ \frac{\text{Tr}\Big[ \hat A_K^\dagger \hat a_j \hat A_K \rho_G \Big]}{\text{Tr}\Big[ \hat A_K^\dagger \hat A_K \rho_G\Big]},
\end{equation}
expanded by Wick--Isserlis theorem in terms of $\rho_G$ statistics $\bm{\hat b}$ and $\Sigma$ from Eq. \eqref{affine-stats} as
\begin{equation}
\begin{split}
\label{qonn-layer-output}
    \langle \hat x_j \rangle_\rho =&\underbrace{\sqrt{2} \Re \big(\bm{\bar b}_j\big)}_{\text{Linear map of }\alpha}+\\
    &\underbrace{\sqrt{2} \Re \Big(\sum_{k\in K} \frac{\bm{\bar b}_k \Sigma_{N+k,N+j} + \bm{\bar b}_{N+k} \Sigma_{k,N+j}}{|\bm{\bar b}_k|^2 + \Sigma_{N+k,N+j}}\Big)}_{\text{Nonlinear map of }\alpha}\ +\\
    &{\mathcal{F}\left(|K| \geq 2\text{ cross-terms}\right)}
\end{split}
\end{equation}

\begin{table*}[t]
    \centering
    \renewcommand{\arraystretch}{1.2}
    \setlength{\tabcolsep}{6pt}
    \begin{tabular}{l l l l}
        \hline
        \textbf{Classical NN} & \textbf{CV-QONN} & \textbf{QONN Free parameters}  & \textbf{Photonic hardware} \\
        \hline
        Inputs $\bm\alpha$
        & $\displaystyle \hat D_{\bm \alpha}$ 
        & N/A
        & BS$(\eta \ll 1)$ + LO \\
        
        Weights $W$
        & \textit{$\hat Q = \hat U_2 \hat S \hat U_1$}
        & $2N^2 + N =
        \begin{cases}
            N^2 \text{ per } \hat U \\
            N \text{ for } \hat S
        \end{cases}$
        & $\begin{cases}
             U \equiv \text{PS + BS}\\
            Z \equiv \hat S = \bigotimes_{j=1}^N \hat S_j
        \end{cases}$\\
        
        Bias $\bm d$
        &$\displaystyle \hat D_{\bm \delta}$
        & $2N$ ($\bm \delta \in \mathbb{C}^N$) 
        & BS$(\eta \ll 1)$ + LO \\
        
        Activation $\Phi$
        & $\hat A_K$
        & $\hat G$ parameters
        & Ancilla mode + BS$(\eta \ll 1)$ + SPD \\
        
        Neurons
        & $|K|$
        & $\hat G$ parameters
        & Ancilla mode + BS$(\eta \ll 1)$ + SPD \\
        
        Outputs $j$
        & $\langle\hat x_{j}\rangle$ 
        & $1$ ($\theta$ phase -- Optional)
        & BS$(\eta = 0.5)$ + LO + PS$(\omega)$ + 2 PD\\
        \hline
    \end{tabular}
    \caption{Relation of standard NN components with quantum optics operators along with their degrees of freedom and their hardware realization, constituting the QONN in Fig. \ref{fig:physical-qnn}.}
    \label{tab:nn-qnn-components}
\end{table*}

This closed-form establishes the evidence that photon subtractions in distinct optical modes of the system give rise to a linear combination of single-ridge activations---effectively one quantum-optical neuron per photon-subtracted mode---so a single-layer QONN with linear readout achieves the UAT on compact sets \cite{cybenko1989approximation, leshno1993multilayer}. When two or more photon-subtracted modes are correlated by non-zero off-diagonal covariances induced by the passive mixer $U_2$ after the squeezing, additional nonlinear cross terms appear enriching the quantum neuron operation, thus increasing the QONN expressivity and potentially reducing the width $|K|$ needed to reach a target accuracy.

According to Eq. \eqref{num_lpms}, the total number of independent terms in Eq. \eqref{qonn-layer-output} grows combinatorially with the amount of distinct loop perfect matchings in the ladder polynomial, this is, with the number of subtracted modes correlated right before the subtraction operations. Thus, for $|K|=1$ the number of terms is $4$ as seen in Eq. \eqref{ph-sub-expval}, while in Eq. \eqref{qonn-layer-output} for $|K|=2$ there are $26$ terms as expanded in appendix \ref{app:two-mode-subtraction}, giving place to highly-correlated quantum optical neurons.

Similarly to Fig. \ref{fig:physical-qnn}, table \ref{tab:nn-qnn-components} shows how each NN component is mapped to a quantum optical component that imitates its operation within the photonic formalism, indicating the number of free parameters of each quantum element. Generalizing for an $L$-layer QONN of $N$ modes, the total number of parameters to be optimized equals
\begin{equation}
\label{qnn-params}
    L(2N^2 + 3N).
\end{equation}

\subsection{Practical implications}
Although the introduced QONN model is adequately described and can be physically realized with current technology, some concerns about the design implementation need to be addressed.

First, the fact that the amounts of QONN inputs, outputs and neurons are directly tied to the available modes imposes the condition of having as many optical modes as the maximum of desired inputs/outputs/neurons quantities.

With respect to hardware limitations, all the quantum optical elements employed are physically available except the strength of the squeezing operator, which is currently restricted to certain ranges when it comes to its realization in the laboratory \cite{schnabel2017squeezed}. Recent works as \cite{vahlbruch2024squeezinglims} reported a ceiling of $15$ dB $ \approx |r| = 1.73$ for the squeezing factor. Although the optimal nonlinear range for $\alpha$ seen in Eq. \eqref{alpha-nonlinear-range} is constrained by the squeezing strength $r$ and by the lower-bound of nonlinear contribution $\varepsilon$, the trainable affine mapping from Eq. \eqref{affine-stats} would always bring the uploaded inputs $\bm \alpha$ to a convenient regime, where the nonlinearities of the subtractions have a non-trivial activation effect on the optical quadratures, avoiding the necessity of setting an explicit range for $\bm \alpha$.

Lastly, the physical implementation of the QONN is intrinsically probabilistic and conditioned on the success probability of the heralded photon-subtraction events. The hardware realization of the annihilation operator in a given signal mode consists in mixing the target mode $\hat a_k$ with an ancillary vacuum mode $\hat b$ on a weak tap beam splitter of signal reflectivity $\eta \ll 1$, followed by single-photon detection on the ancilla arm \cite{ourjoumtsev2006generating,dakna1997generating,wenger2004nonGaussian,garcia2004proposal}. The corresponding two-mode beam splitter unitary acts as
\begin{equation}
    \hat U_{\mathrm{BS}}:\begin{pmatrix}
        \hat a_k\\
        \hat a_\text{anc}
    \end{pmatrix}
    \mapsto 
    \begin{pmatrix}
        \sqrt{1-\eta}\,\hat a_k+\sqrt{\eta}\,\hat a_\text{anc}\\
        \sqrt{1-\eta}\,\hat a_\text{anc}-\sqrt{\eta}\,\hat a_k
    \end{pmatrix},\ [ \hat a_k,\hat a_\text{anc}^\dagger ] = 0.
\end{equation}
Conditioning on a single click on the ancilla mode and expanding to first order in $\eta \ll 1$, the effective Kraus operator acting on the signal is
\begin{equation}
    \hat M = {}_{\mathrm{anc}}\!\langle 1|\,\hat U_{\mathrm{BS}}\,|0\rangle_{\mathrm{anc}}
    \ \propto\ \sqrt{\eta}\,\hat a_k,
\end{equation}
which, applied to the final Gaussian state $\rho_G$ from Eq. \eqref{affine-stats}, brings the system to the unnormalized conditional state $\tilde\rho'=\hat M\,\rho_G\,\hat M^\dagger\propto \hat a_k\rho_G\hat a_k^\dagger$. Therefore, the single-subtraction success probability is 
\begin{equation}
\label{subtraction-success}
    p_{\mathrm{succ}}^{(1)}=\mathrm{Tr}[\tilde\rho']
    \approx \eta\,\mathrm{Tr}[\hat a_k^\dagger\hat a_k\rho_G]
    = \eta\,\big(\Sigma_{N+k,N+k} + |\bm{\bar b}_k|^2\big)
\end{equation}

For simultaneous subtractions in a set of distinct modes $K$ with independent reflectivities $\eta_k$ and SPDs, the joint Kraus operator is $\hat M_K\propto \prod_{k\in K}\sqrt{\eta_k}\,\hat a_k$, giving
\begin{equation}
\label{multi-subtraction-success}
    p_{\mathrm{succ}}^{(K)}=\Big(\prod_{k\in K}\eta_k\Big)\,\big\langle \hat A_K^\dagger \hat A_K\big\rangle_{\rho_G},
\end{equation}
which can be evaluated exactly from the first and second order moments of $\rho_G$ by Wick--Isserlis expansion. In the common weak–tap regime $\eta_k\ll 1$ and when multi-photon events are negligible in each tap, a useful approximation is
\[
p_{\mathrm{succ}}^{(K)}\approx \prod_{k\in K}\eta_k\,\langle \hat a_k^\dagger\hat a_k\rangle_{\rho_G}
=\prod_{k\in K}\eta_k\,\big(\Sigma_{N+k,N+k} + |\bm{\bar b}_k|^2\big),
\]
consistent with multi-mode analysis of engineered addition/subtraction sequences \cite{fiurasek2009engineering}. In practice, one may fold detector inefficiency, optical loss and mode-matching into an effective tap $\eta\to\eta_{\mathrm{eff}}$ without changing the form of the expressions above.

\subsection{Generalization to ladder polynomials}
While this architecture focuses on photon subtraction as the elementary source of nonlinearity, the developed formalism is, in fact, much more general. The key ingredient of the construction is that the non-Gaussian layer acts via a polynomial in the ladder operators on an underlying Gaussian state as in Eq. \eqref{perf_match}. Instead of the product of annihilation operators $\prod_{k\in K} \hat{a}_k$ from Eq. \eqref{multi-subtraction-op} used for heralded photon subtraction, one may consider a general operator of the form
\begin{equation}
\label{eq:general_polynomial}
    \hat{\mathcal{O}}_{K} \;=\; \prod_{k \in K} P_k(\hat{a}_k,\hat{a}_k^\dagger),
\end{equation}
where each $P_k$ is a finite polynomial in $\hat{a}_k$ and $\hat{a}_k^\dagger$. The Wick--Isserlis-based machinery presented in this work still provides exact expressions for all homodyne moments of the state $\hat{\mathcal{O}}_{K} \ket{\psi_{\mathrm{G}}}$, and therefore for the effective activation functions induced by such general non-Gaussian layers. Photon subtraction is thus only one particularly convenient and experimentally accessible instance of a much broader class of polynomial non-Gaussian resources.

This observation directly connects the introduced framework to perturbative implementations of weak non-Gaussian gates such as cubic or Kerr interactions \cite{lloyd1999quantum, braunstein2005quantum, yanagimoto2020engineering}. A short-time evolution under a non-Gaussian Hamiltonian $\hat{H}_{\mathrm{ng}} \in \{\hat{H}_{\mathrm{cubic}}, \hat{H}_{\mathrm{Kerr}}, \dots\}$ can be written as a power series
\begin{equation}
    e^{- i t \hat{H}_{\mathrm{ng}}}
    \;=\;
    \mathbb{1} - i t \hat{H}_{\mathrm{ng}}
    - \tfrac{t^2}{2} \hat{H}_{\mathrm{ng}}^2
    + \mathcal{O}(t^3),
\end{equation}
where each term is a polynomial in the bosonic ladder operators. In the weak-gate regime (fast-gate), retaining only the leading orders in $t$ precisely yields the kind of polynomial structure captured by Eq. \eqref{eq:general_polynomial}. The ansatz proposed in this work therefore provides an exact, non-perturbative evaluation of the homodyne statistics associated with such perturbative non-Gaussian gates, and can in particular reproduce and systematize the continuous-variable neural network constructions proposed by Lloyd and co-workers \cite{lloydcvqnn} and by Siopsis and collaborators \cite{siopsis2023exprealcvqnn} within a single unified framework without truncation.

From a hardware perspective, this generality implies that this scheme is not restricted to photonic platforms employing heralded photon subtraction. Whenever a platform can implement weak non-Gaussian interactions whose short-time expansion gives rise to polynomials in ladder operators, the same theoretical machinery and universality results apply. Examples include levitated opto-mechanical systems \cite{bemani2025heralded, muffato2025generation, marti2024quantum}, superconducting microwave circuits \cite{blais2021circuit, kounalakis2018tuneable, hu2023fast}, and other bosonic architectures where cubic, Kerr, or more general non-Gaussian Hamiltonians are available. In this sense, the photon-subtraction implementation studied here should be viewed as a particularly mature and experimentally feasible instance of a broadly applicable continuous-variable quantum optical neural network paradigm.

\section{Benchmarks}
In this section, supervised learning---classical and quantum---tasks such as curve fitting, classification, and complex non-Gaussian gate synthesis are tackled to test the performance of the QONN in different contexts.

For numerical experiments, section \ref{sec:code-avail} describes the features of the \texttt{QuaNNTO} library developed to perform exact classical simulations of the CV-QONN in Fig. \ref{fig:physical-qnn}.

Each simulation of these experiments has been executed in \emph{MareNostrum 5}, the supercomputer of \emph{Barcelona Supercomputing Center}, using a single node of 112 cores from the general-purpose partition.

Given the physical constraint of squeezing strength, all the results are bounded within a state-of-the-art limitation range $r \in (-1.7, 1.7)$ keeping simulation experiments realistic.

\subsection{Curve fitting}
The curve fitting tasks carried out in this section aim to prove the QONN learning capacity and performance in regression-like (CV) problems. For such tasks, the goal is to train a NN intending to learn a continuous function fringe out of some---noisy---samples of it, which build up the dataset.

For the sake of diversity and generality, multiple complex functions are chosen to test the UAT in the QONN architecture. To do so, training benchmarks are performed for such functions on different single-layer QONN setups as purely Gaussian, single-neuron and multi-neuron.

\begin{figure*}[ht]
    \begin{subcaptionblock}[b]
    {0.33\textwidth}
        \includegraphics[width=\linewidth]{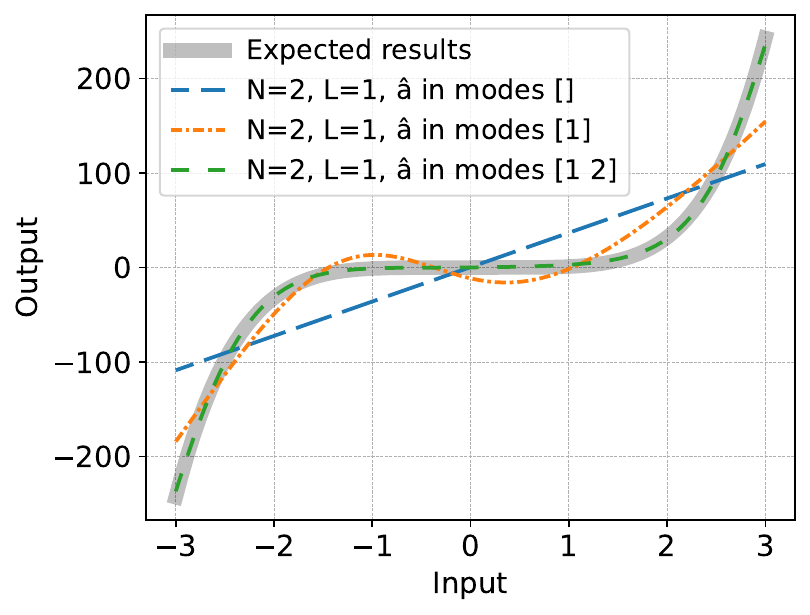}
    \end{subcaptionblock}
    \begin{subcaptionblock}[b]
    {0.33\textwidth}
        \includegraphics[width=\linewidth]{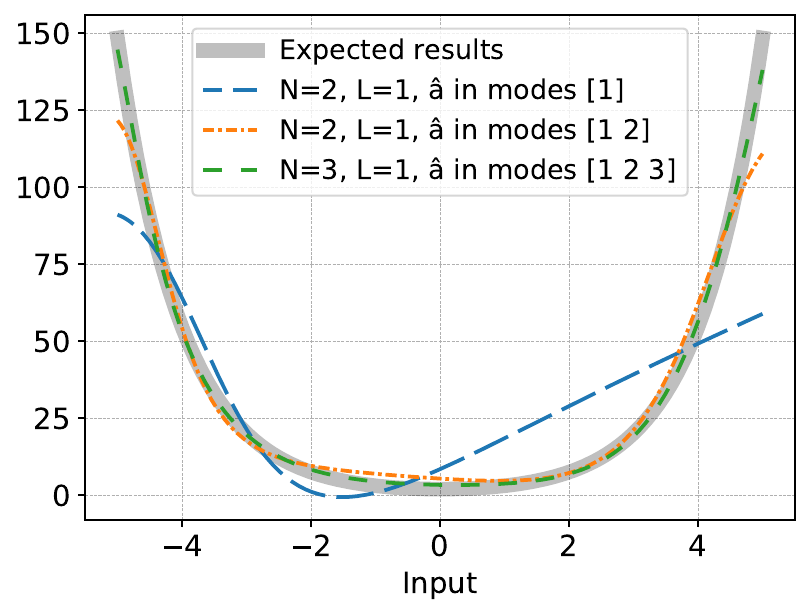}
    \end{subcaptionblock}
    \begin{subcaptionblock}[b]
    {0.33\textwidth}
        \includegraphics[width=\linewidth]{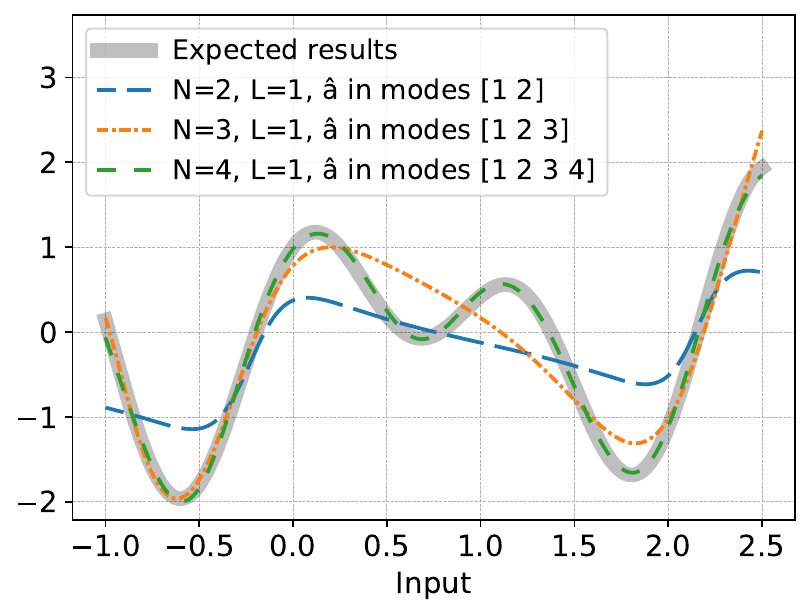}
    \end{subcaptionblock}
    \caption{Curve fitting training of $N$-mode single-layer ($L=1$) QONNs with photon subtraction $\hat a$ in different modes trained over 100-sample noisy datasets of exponential, polynomial and oscillatory functions. Left: Gaussian, single-neuron and two-neuron QONNs results for $f(x) = x^5$ with $x \in (-3, 3),\ \epsilon = 5$. Center: one, two and three-neuron QONNs results for $f(x) = 2\cosh(x)$ with $x \in (-5, 5),\ \epsilon = 3$. Right: two, three and four-neuron QONNs results for $f(x) = \sin (3x) + \cos (5x)$ with $x \in (-1, 2.5),\ \epsilon = 0.1$.}
    \label{fig:curve-fitting}
\end{figure*}

All simulation experiments have been performed with the following specifications:
\begin{itemize}
    \item Synthetic training dataset of a target function $f(x)$ constructed by randomly sampling 100 points within a specific range of $x$ from a uniform distribution with some noise level $\epsilon$.
    \item Validation set made of 50 noiseless and random samples of $f(x)$ in the specified range of $x$.
    \item Test set made of 200 noiseless and linear samples of $f(x)$ in the specified range of $x$.
    \item Output: Expectation value of first-mode position observable $\langle \hat x_1 \rangle$ acting as the QONN predicted outcome $\tilde f(x)$---like in Eq. \eqref{exp_val-single-layer-example}.
    \item Pre-processing: re-scaling of inputs $x$ to range of displacement $\alpha_1 \in (-3,3)$, and re-scaling of outputs $f(x)$ to $\langle \hat x_1 \rangle \in (1, 5)$.
    \item Mean-Squared Error (MSE) as loss function aiming to reduce the distance between the predicted and the actual values in the dataset
        \begin{equation}
        \label{mse}
            \text{MSE } = \frac{1}{M} \sum_{j=1}^M(f(x_j) - \tilde f(x_j))^2,
        \end{equation}
        where $M$ is the number of samples in the dataset, $f(x_j)$ is the desired output value for the input $x_j$, and $\tilde f(x_j) = \langle \hat{x}_1 \rangle_{x_j}$ is the QONN predicted value for such input.
\end{itemize}

Fig. \ref{fig:curve-fitting} shows the different single-layer QONN trainings indicating the modes where a photon subtraction is present. The left plot proves that, without subtractions---i.e. QONN with only Gaussian operators---, learning is limited to a linear fit due to the lack of activations. Overall, these results show that adding more neurons---arising from the amount of subtracted modes---in a single layer makes the fitting error tend to zero regardless the complexity of the function, in line with the UAT.

Additionally to the functions of Fig. \ref{fig:curve-fitting}, the QONN reports perfect training---similar to the displayed ones---for rational, hyperbolic tangent, Gaussian, absolute value, and logarithmic functions, whose results can be obtained through the \texttt{QuaNNTO} library available in section \ref{sec:code-avail}.

\subsection{Classification}
\label{sec:classification}
Supervised classification tasks are another main focus of NNs, where the network's aim is to create relations among the inputs or features in order to establish the category they belong to.

Considering that classification tasks involve discrete-domain datasets, the data must be adapted to make such tasks suitable for the CV formalism. This involves the treatment of the QONN continuous outcomes in order to interpret them as classification labels.

Explicitly, the classification experiments of this section are carried out with the following general characteristics:
\begin{itemize}
    \item Outputs: one observable per category, $\langle \hat x_j \rangle$, where $j$ denotes the optical mode and the category. This is, for $C$ different classes, $\langle \bm{\hat x} \rangle = (\langle \hat x_1 \rangle,...,\langle \hat x_C \rangle)$.
    \item Pre-processing: re-scaling of the inputs to range of displacement $\alpha_j \in (-3, 3)$. Re-scaling of the outputs to range $\langle \hat x_j \rangle \in (1, 5)$, where each entry is represented by a \emph{one-hot encoding} vector---a processing technique to go from categorical to numerical variables---denoting the belonging class of an input.\\
    E.g.: for a dataset entry that belongs to class number \emph{2} out of \emph{4} existing classes, its re-scaled \emph{one-hot} output is $\langle \bm{\hat x} \rangle = (1, 5, 1, 1)$.
    \item Post-processing: Softmax discretization to transform the continuous quantities $\langle \hat x_j \rangle$ into probabilities like
        \begin{equation}
        \label{softmax-disc}
            p_j = \frac{e^{z_j}}{\sum_{k=1}^{C} e^{z_k}},
        \end{equation}
        where $z_j$ is the raw outcome of $\langle \hat x_j \rangle$, $C$ is the number of classes and $p_j$ is the probability of belonging to class $j$.\\
        The highest probability $p_j$ obtained for a given dataset entry will determine its predicted belonging category.
        
    \item Cross-entropy loss function over the post-processed probabilities is used to measure the difference between the predicted probability distribution and the true distribution. It is defined as
        \begin{equation}
        \label{cross-entropy}
            \text{Cross-Entropy} = - \frac{1}{M} \sum_{i=1}^{M} \sum_{j=1}^{C} y_{i,j} \log(p_{i,j}),
        \end{equation}
        where $M$ is the number of samples, $C$ is the number of categories, $y_{i,j} \in \{0,1\}$ indicates whether the sample $i$ belongs to class $j$, and $p_{i,j}$ is the predicted probability of sample $i$ belonging to class $j$.
\end{itemize}

Given these classification features, the QONN training will consist of---in physical terms---maximally displacing the optical mode denoting the correct category of a given input. In other words, the larger the position quadrature mean of a given mode is, the higher the probability of belonging to such a class.

\subsubsection{Moons and circles classification}
To test the classification skills of the QONN model, two toy-example tasks to be tackled are the 2D \emph{moons} and \emph{circles} classification. In these problems, the target of the network is to learn a nonlinear boundary that tells apart two different classes of points arranged in a continuous two-dimensional grid.

These datasets can be generated through \textit{scikit-learn} Python's library by setting different samples and noise levels. Specifically, the employed datasets for these benchmarks contain 500 balanced samples of noise $\epsilon=0.1$ for \emph{moons} and $\epsilon=0.05$ for \emph{circles}. The training sets are composed of 100 balanced samples---randomly chosen from the dataset---(50 for the validation sets), and the trained model is finally tested with all 500 samples.

According to the previously defined features, QONNs of $N=2$ modes are sufficient to perform these tasks, since both problems are binary classifications and the input space is two-dimensional and real-valued.

\begin{figure}[ht]
    \begin{subcaptionblock}[b]{0.24\textwidth}
        \includegraphics[width=\textwidth]{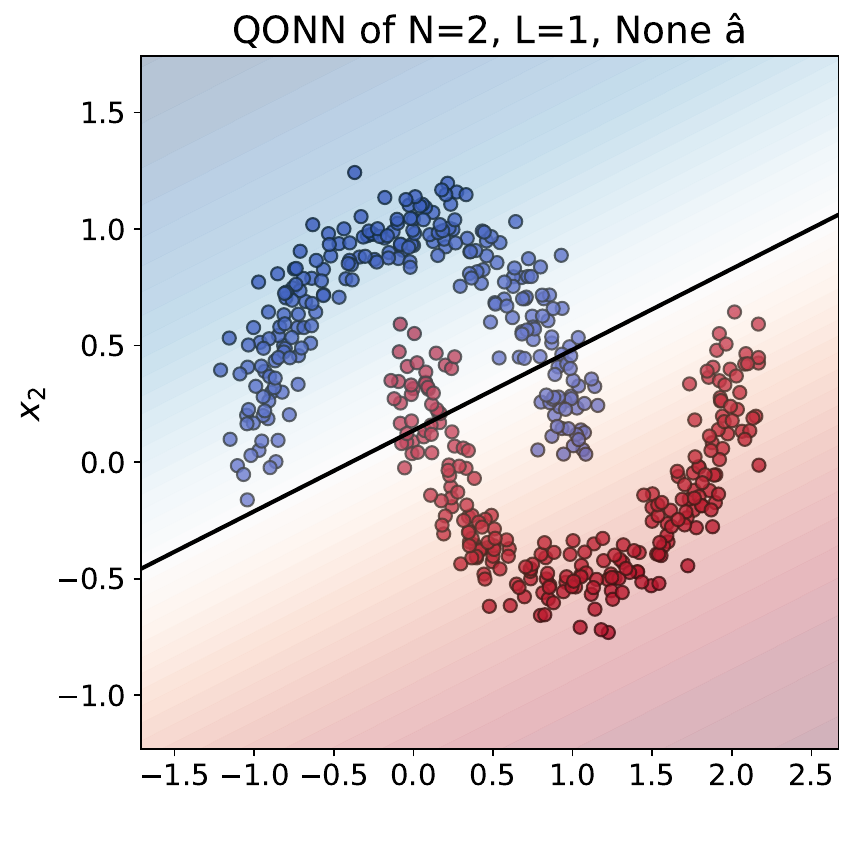}
    \end{subcaptionblock}
    \centering
    \begin{subcaptionblock}[b]{0.24\textwidth}
        \includegraphics[width=\textwidth]{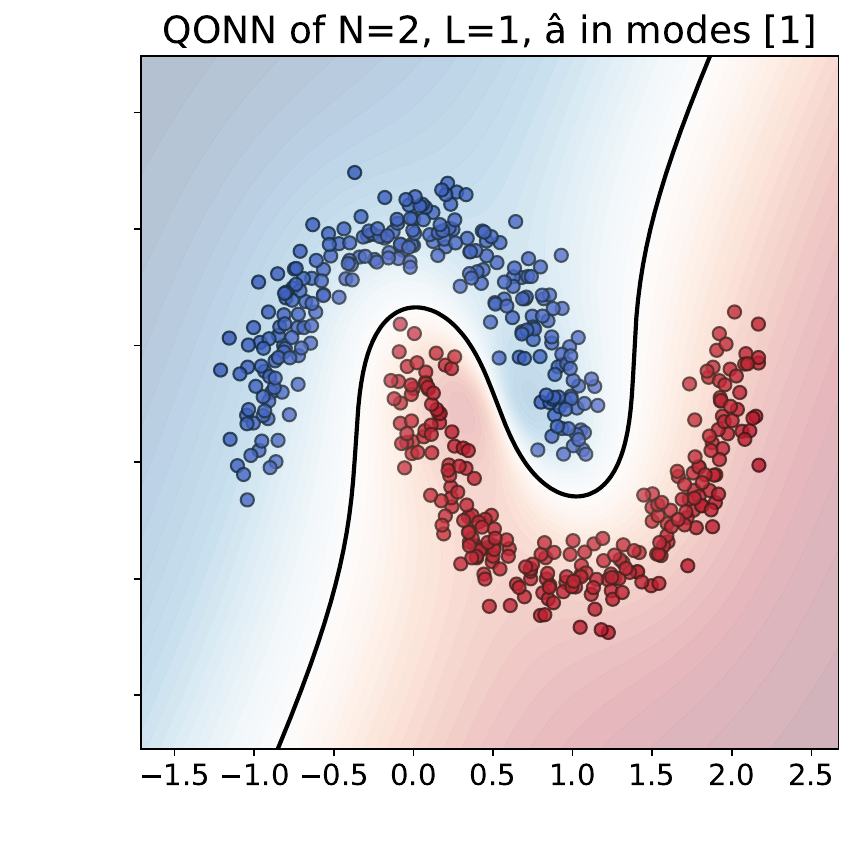}
    \end{subcaptionblock}
    \vspace{1mm}
    \begin{subcaptionblock}[b]{0.24\textwidth}
        \includegraphics[width=\linewidth]{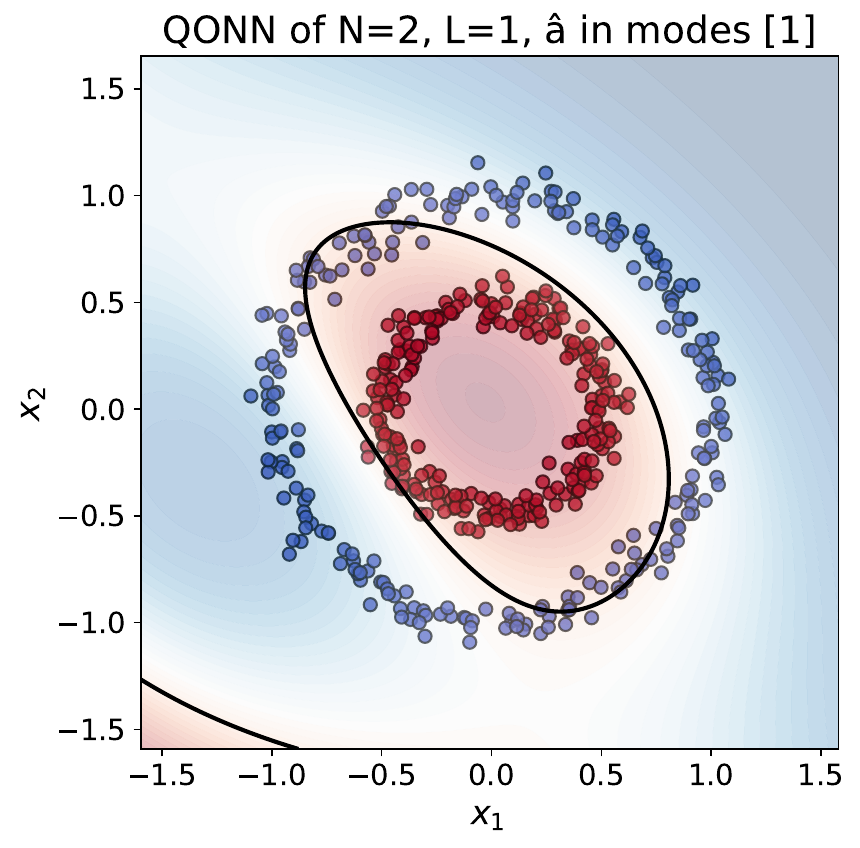}
    \end{subcaptionblock}
    \centering
    \begin{subcaptionblock}[b]{0.24\textwidth}
        \includegraphics[width=\linewidth]{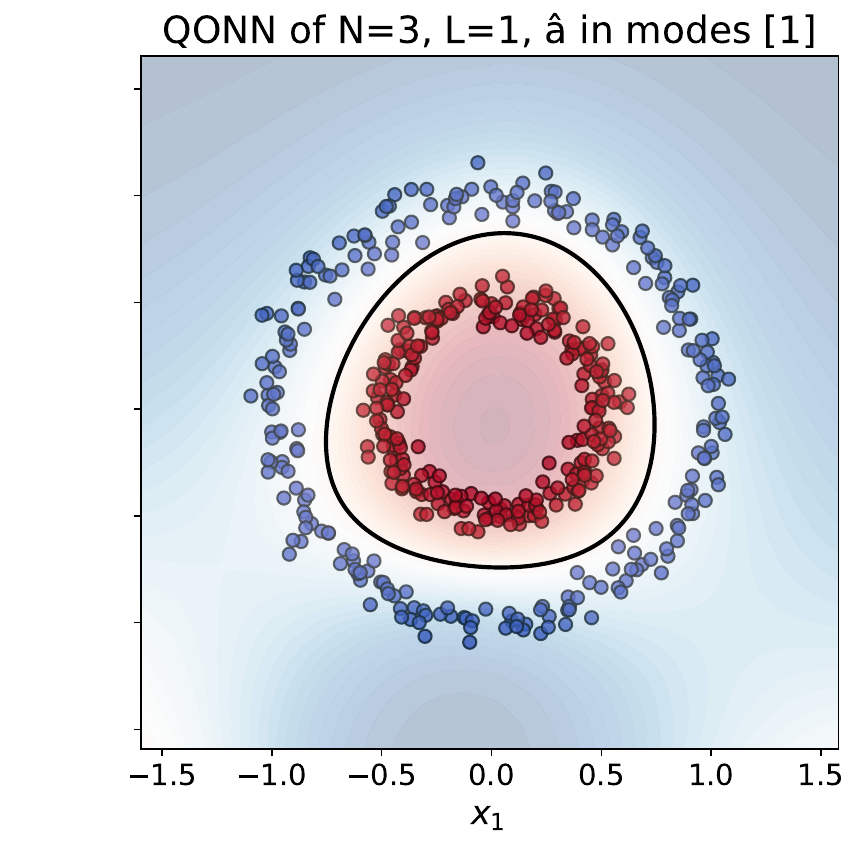}
    \end{subcaptionblock}
    \caption{\emph{Moons} classification (upper row): two interleaved categories of points. Upper-left: Linear decision boundary with 86.6\% accuracy of a Gaussian QONN (no subtractions). Upper-right: S-like decision boundary with 100\% of accuracy of a single-layer single-subtraction QONN.
    \emph{Circles} classification (lower row): two concentric categories of points. Lower-left: ellipse-like decision boundary with 91.4\% of accuracy of a single-layer single-subtraction QONN of $2$ modes. Lower-right: radial decision boundary with 100\% of accuracy of a single-layer single-subtraction QONN of $3$ modes.} 
    \label{fig:moons_circles_class}
\end{figure}

In Fig. \ref{fig:moons_circles_class}, the upper row shows how a Gaussian QONN trains up to a linear decision boundary for the \emph{moons} problem, while a single-layer QONN with one neuron is able to fully discern between the two classes with a curved decision boundary, achieving perfect accuracy. On the other hand, the \emph{circles} problem needs a radial decision boundary in order to be solved. The lower row of Fig. \ref{fig:moons_circles_class} demonstrates that a single-neuron QONN with two modes fails to capture such a boundary, where a single-neuron QONN with 3 modes instead succeeds with perfect precision. In this last case, adding the third optical mode gives freedom and balance to the output probabilities during the training of the two-category classification since one of the system modes is not observed, that is, the third mode acts as an auxiliary mode that modulates the outcomes of the first two modes whose output values are the ones denoting the probability of belonging to their respective classes.

\subsubsection{MNIST classification}
As a practical classification task with discrete inputs, an experiment of multi-category grouping is provided using the famous MNIST dataset---$28\times28$ pixels images of handwritten numbers from 0 to 9---where a subset of categories is selected.

In this particular case, the inputs would be the discrete values of a matrix that represent the pixels of an arbitrary dataset image. Then, one way of adapting the dataset for the CV framework is using \emph{Principal Component Analysis} (PCA) or \emph{Autoencoders} over the images to obtain a real-valued vector that encodes each image, allowing its uploading in the initial coherent states.

Additionally to the general features of classification tasks defined in section \ref{sec:classification}, the MNIST task is performed considering:
\begin{itemize}
    \item Five different labels or categories: $\{0, 1, 2, 3, 4\}$.
    \item Real-valued inputs: Autoencoder compressing the $28\times28$ images to a continuous latent space of $5$ real-domain dimensions.
    \item QONN of $N=5$ modes, one per category, and $2$ inputs corresponding to the latent space dimension of the images.
    \item Training set of 75 random images per category, a total of 375 entries in this case.
    \item Validation set of 20 images per category, making a total of 100 images.
    \item Test set with all images from the dataset ($\sim$ 5000).
\end{itemize}

\begin{figure}[ht!]
    \centering
    \begin{subcaptionblock}[b]
    {0.48\textwidth}
        \includegraphics[width=\textwidth]{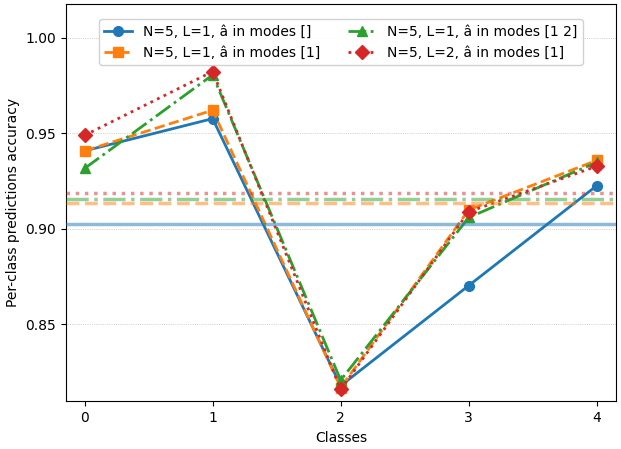}
    \end{subcaptionblock}
    \centering
    \caption{5-category MNIST classification showing per-class accuracy (opaque lines) and average precision (transparent horizontal lines) for different QONN models with an input pre-processing of 5-dimensional latent space.}
    \label{fig:mnist-classification}
\end{figure}

Fig. \ref{fig:mnist-classification} shows that adding more neurons in a single-layer QONN enhances the classification capacity, and that, with the same non-Gaussian resources, a two-layer QONN with one subtraction per layer slightly outperforms the two-neuron single-layer model.

Considering that the MNIST problem has a purely discrete dataset, these results show that there are functional data pre-processing alternatives to adjust discrete-variable problems to the CV environment where the QONN would remain effective in solving such problems.

\subsection{Non-Gaussian gate synthesis}
\begin{figure*}[ht]
    \begin{subcaptionblock}[b]{\columnwidth}
        \includegraphics[width=\linewidth]{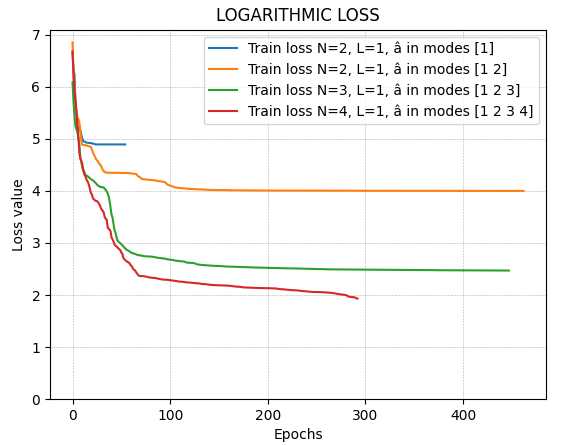}
    \end{subcaptionblock}
    \centering
    \begin{subcaptionblock}[b]{\columnwidth}
        \includegraphics[width=0.95\linewidth]{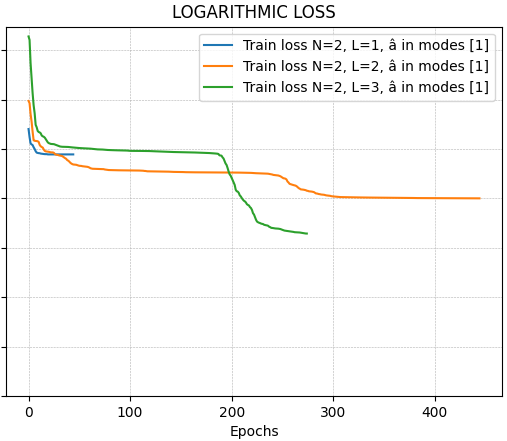}
    \end{subcaptionblock}
    \caption{Cubic-phase non-Gaussian gate synthesis where different QONNs of Fig. \ref{fig:physical-qnn} learn the set of statistical moments of the quantum states $\ket{\psi(\alpha_j)} = V(\gamma)\ket{\alpha_j}$ with $\alpha \in (-2,2)$ and $\hat{V}(\gamma)=e^{i\gamma \hat{x}^3/3}$ of strength $\gamma = 0.2$. Left: Losses of the statistical infidelity using single-layer QONNs with one, two, three and four neurons. Right: Losses of the statistics infidelity using QONNs with one, two and three layers.}
    \label{fig:nongauss-gate-synth}
\end{figure*}

The QONN simulation technique used in this work allows the computation of expectation values of any observable over any non-Gaussian state as long as the final state's wavefunction can be written as ladder operators acting over a Gaussian state $\rho_G$.

This admits to train the proposed QONN to match the statistical moments of any complex non-Gaussian gate acting over a particular set of quantum states. This means that the trained QONN block would be able to replicate the action of the chosen non-Gaussian operator upon the target quantum states using physically available quantum optical components, similar to \cite{yukawa2013emulating}.

As a demonstration, the target is to synthesize the action of the cubic phase operator $\hat{V}(\gamma)=e^{i\gamma \hat{x}^3/3}$ over coherent states $\ket{\alpha}$ in the range of $\alpha \in (-2,2)$. To do so, the Strawberry Fields platform \cite{killoran2019strawberry} is used to build up the dataset for the QONN with the following specifications:
\begin{itemize}
    \item $M$ linearly separated samples of real displacements $\alpha \in (-2,2)$ defining the input set of coherent states ${\ket{\alpha_j}}$ with $j \in \{1,...,M\}$. These real coherent states are the input-encoded states of the QONN.
    \item Action of the operator over the different coherent states $\ket{\psi(\alpha_j)} = V(\gamma)\ket{\alpha_j}$ for a fixed $\gamma$.
    \item For each final quantum state $\ket{\psi(\alpha_j)}$, compute all statistical moments up to fourth-order $O=\{\langle \hat a \rangle,\langle \hat a^2 \rangle,\langle \hat N \rangle,\langle \hat a^3 \rangle,\langle \hat a^2\hat a^\dagger \rangle, \langle \hat N^2 \rangle\}$ with $\hat N = \hat a^\dagger \hat a$. This set constitutes the outputs of the dataset.
\end{itemize}

The loss function for this task is, again, MSE but for all moments in the set like
\begin{equation}
\label{mse_set}
    \text{MSE } = \frac{1}{M} \sum_{\hat o \in O}\sum_{j=1}^M(\langle \hat o \rangle_{\alpha_j} - \langle \tilde{\hat{o}} \rangle_{\alpha_j})^2,
\end{equation}
where $\langle \hat o \rangle_{\alpha_j}$ is the actual value of the statistical moment $\hat o$ of $\ket{\psi(\alpha_j)}$ and $\langle \tilde{\hat{o}} \rangle_{\alpha_j}$ is the QONN predicted value for $\langle \hat o \rangle_{\alpha_j}$.

From Fig. \ref{fig:nongauss-gate-synth}, the achieved losses in the left plot demonstrate how the precision of such gate synthesis is improved within a single-layer QONN when the number of neurons increases. The same effect is observed in the right losses where the synthesis is trained with single-subtraction QONNs of multiple layers, boosting its accuracy when more layers are added. The performance contrast between multi-neuron and multi-layer QONNs with an identical amount of non-Gaussian resources shows some symptoms of vanishing gradients in the multi-layer scenario since, as in the case of a single-layer QONN with two subtractions and a two-layer QONN with one subtraction per layer (orange curves), both loss convergences tend to the same value, taking more epochs for the two-layer instance, which manifests a local minima clog until approximately the 250th epoch, while in the single-layer case such a convergence value is reached in around 100 epochs. This is also supported by the cases of three subtractions (green curves) which show that a three-layer model gets easily stuck in local minima and it did not get to converge to a value close to the one achieved by the single-layer model with three subtractions.

It is worth mentioning that the synthesis of $V(\gamma)$ operator could be performed accounting $\gamma$ as another parameter to be learned instead of fixing its value. Similarly, the process could be extended to other target quantum states, such as general Gaussian states, which would require more operators and parameters to be encoded in the QONN. Furthermore, the same synthesis process could be applied to other complex non-Gaussian operators like the Kerr gate $\hat{K}(\kappa) = e^{i\kappa\hat{N}^2}$.

Note that, for accuracy purposes, the finite set $O$ can be extended by increasing the order of statistical moments considered in the QONN training, depending on the complexity of the operator to be replicated. Unlike minimizing the infidelity of the QONN with the gate action, this approach of matching a finite set of statistical moments is realistic in an experimental setting.

\section{Discussion}
The CV-QONN model proposed in Fig. \eqref{fig:physical-qnn} based on Gaussian operators and multi-mode subtractions sequences unifies experimental feasibility, saving non-Gaussian resources, analytic tractability, and expressive power in a single building block for CV-QONNs. The Wick--Isserlis-based simulation framework---plus Bogoliubov transformations for multi-layer setups---enables exact computation of expectation values and gradient evaluation in high-performance environments without Hilbert space truncation, coordinating both classical training and error analysis, and providing a new efficient ansatz for exact classical simulation.

The presented architecture provides the well-characterized activation function of Eq. \eqref{ph-sub-expval} that leverages the nonlinearity induced by the photon subtraction action on squeezed coherent states, avoiding the need for demanding or challenging non-Gaussian interactions. Furthermore, the UAT is accomplished within a single layer, where each subtracted optical mode builds a quantum neuron and, when two or more squeezed modes are correlated and then photon-subtracted as in Eq. \eqref{qonn-layer-output}, highly-correlated quantum neurons arise involving a stronger expressivity of the model.

As shown in Eq. \eqref{subtraction-success}, the success probability of the subtraction operation is subordinated to the squeezing and displacement magnitudes of the target mode, but can be arbitrarily increased through the reflectivity $\eta$ of the weak-tap BS. Moreover, Eqs. \eqref{sq-inf} and \eqref{sq-minusinf} prove that non-trivial nonlinearity can be perfectly achieved with a small squeezing strength, fitting the state-of-the-art of available optical hardware and ensuring performance of the QONN on low-energy quantum systems.

Benchmarking across diverse tasks---ranging from function approximation and supervised classification on nontrivial datasets to the gate synthesis of complex non-Gaussian operators by matching statistical moments---confirms that single-layer QONNs with enough neurons and single-neuron QONNs with multiple layers achieve high fidelity and robust generalization.

An important conceptual difference between this QONN proposal and earlier continuous-variable neural network architectures \cite{lloydcvqnn,siopsis2023exprealcvqnn} is the role of Fock-space truncation. In those works, all quantities of interest are ultimately computed in a truncated Fock basis. Formally, this renders the scheme equivalent to a---very large---discrete-variable model: the dynamics is projected onto a finite-dimensional subspace, classical simulation inherits the unfavorable scaling of the truncated Hilbert space, and the resulting figures of merit are not directly expressed in terms of experimentally accessible quadrature statistics. In contrast, the presented approach is based on a genuinely continuous-variable ansatz and an \emph{exact} evaluation of homodyne moments, without any photon-number cutoff, by combining Bogoliubov transformations with generalized Wick--Isserlis theorems. This provides a closed-form description of the effective activation functions and a classical simulation framework whose cost scales with the number of modes, layers, and photon subtractions, rather than with an artificially imposed Fock cutoff. For the tasks studied here, substantial practical speedup is observed over state-of-the-art Fock-space simulators such as QuTiP and Strawberry Fields.

These differences have direct implications for both expressivity and experimental feasibility. For instance, in the architecture of \cite{siopsis2023exprealcvqnn}, the network depth reaches up to twelve layers with three photon-subtraction events per layer. At realistic squeezing levels and detection efficiencies, such a configuration would lead to heralding rates that are exceedingly low---on the order of one successful detection per day---, implying training times of years even under optimistic assumptions. To keep the numerics tractable, the analysis therefore relies on a photon-number cutoff, typically $n_{\max}=6$, reintroducing the drawbacks of truncation at the simulation level. In the exact framework introduced, by contrast, a \emph{single} layer with only a few photon-subtracted modes is already sufficient to match or surpass the performance reported \cite{siopsis2023exprealcvqnn} on comparable tasks, while remaining compatible with realistic heralding probabilities on current photonic platforms. This highlights that a shallow, analytically tractable continuous-variable quantum optical neural network can simultaneously achieve high expressive power, efficient classical simulability, and genuine experimental viability without resorting to Fock space truncation.

Taken together, these advantages position the proposed layer as a compelling candidate for scalable photonic quantum machine learning and for quantum applications such as state engineering and non-Gaussian gate synthesis.

Moreover, because the present analytic framework operates exactly for quantum states made of Gaussian operators and finite polynomials in ladder operators, it extends to any CV platform that admits such a description. This opens the door to closed-form derivations of layer activations for a broad class of non-Gaussian elements and measurement schemes. In particular, weak Kerr interactions can be incorporated \emph{perturbatively} via a truncated expansion of the gate to yield analytically tractable, series-defined activation functions suitable for quantum machine learning and CV-QONN architectures.

Future work will focus on improving the \texttt{QuaNNTO} simulator to enhance efficiency and sketch larger QONNs, experimentally implementing and testing the physical setup of the CV-QONN in a quantum optical computer, extending the analytic mechanism to other frameworks, as well as researching for more quantum applications of such a universal quantum design as quantum tomography, non-Gaussian entanglement witnesses and metrology, and quantum field theory.

\section{Code availability}
\label{sec:code-avail}
The specific-purpose Python library \href{https://github.com/bsc-quantic/QuaNNto}{\texttt{QuaNNTO}} has been developed to emulate the QONN in Fig. \ref{fig:physical-qnn} with the exact ladder-polynomial non-Gaussianity simulation technique based on the Bogoliubov unitary action over ladder operators and the non-zero means Wick--Isserlis expansion, respectively supported in appendices \ref{appendix:classical-sim} and \ref{app:two-mode-subtraction}.

Since the QONN outputs are computed as sums of traces, and each of these traces can be evaluated independently via the Wick--Isserlis expansion of Eq. \eqref{perf_match}---which yields a term-wise sum of products---high-performance computing environments are natural candidates for the classical simulation of the model. For this reason, the \texttt{QuaNNTO} library is implemented on top of the high-performance \emph{JAX} framework \cite{jax2018github}, together with the \emph{Basinhopping} method from the \emph{SciPy} library, which internally uses the \emph{L--BFGS--B} optimization algorithm with box constraints. In practice, this realizes a hybrid training strategy: an automatic-differentiation (AD) based warm-up stage, implemented purely in JAX, followed by a quasi-Newton optimization method (L--BFGS--B) acting as a local refiner. This combination enables large-scale simulations of gradient-based QONN training procedures on distributed and parallel classical hardware.

\section*{Authors contribution}
T.K.-I. developed the main research ideas, determined the experimental procedures, coded and run the benchmarks and simulations. A.C.-L. supervised the parts related with the universality of the method and suggested improvements from a quantum machine learning perspective. P. S. and F. C. proposed the main ideas and supervised the work together. All authors reviewed and edited the manuscript.

\section*{Acknowledgements}
P. S. and T. K.-I. acknowledge funding from the Spanish Ministry for Digital Transformation and the Civil Service of the Spanish Government through the QUANTUM ENIA project call - Quantum Spain, EU, through the Recovery, Transformation and Resilience Plan – NextGenerationEU, within the framework of Digital Spain 2026. A.C.-L. acknowledges funding from Grant RYC2022-037769-I funded by MICIU/ AEI/ 10.13039/ 501100011033 and by “ESF+" and from PID2023-147245NA-I00 (LOGITONICS) project funded by the MCIU /AEI /10.13039/501100011033 / FEDER, UE. F.C. acknowledges funding from the European Union (EQC, 101149233).

\bibliographystyle{unsrt}
\bibliography{sample}

\onecolumn

\numberwithin{equation}{section}
\renewcommand{\theequation}{\thesection\arabic{equation}}
\appendix
\section*{Appendix}

\section{Nonlinearity of the creation operator in phase-space}
\label{app:photon-addition-nonlinearity}
The photon addition operator $\hat a^\dagger$ also has a nonlinear effect in phase-space. Unlike photon subtraction, $\hat a^\dagger$ does not need a previously-squeezed state to induce nonlinearity on the quantum---coherent---state encoding the input.

Here, two different setups of the photon addition operator for nonlinearity injection in the $\hat x$ quadrature are analyzed: the case of multiple additions acting on real coherent states and a photon addition acting on a $r$-squeezed real coherent states.

First, let's consider the single-mode non-Gaussian state 
\begin{equation}
\label{ph-add-coherent-wf}
\ket{\psi_n(\alpha)} = (\hat a^\dagger)^{n}\,\hat D(\alpha)\ket{0} \text{ with } \alpha \in \mathbb{R},
\end{equation}
where a displacement operator followed by $n$ photon-addition operators act on the vacuum state. The normalized expectation value
\begin{equation}
\label{ph-add-expval}
\langle \hat x(\alpha) \rangle_n = \frac{\bra{\psi_n(\alpha)} \hat x \ket{\psi_n(\alpha)}}{\langle{\psi_n(\alpha)}\ket{\psi_n(\alpha)}} = \frac{
 \bra{0} \hat D^\dagger(\alpha)\,\hat a^{n} \ \hat x\,(\hat a^\dagger)^{n}\,\hat D(\alpha)\ket{0}
}{
 \;\bra{0} \hat D^\dagger(\alpha)\,\hat a^{n}\,(\hat a^\dagger)^{n}\,\hat D(\alpha)\ket{0}
}\,
\end{equation}
will measure the amount of nonlinearity produced by the photon-addition operators on different input states $\alpha$.

Eq. \eqref{ph-add-expval} is equivalent to Laguerre polynomials defined as creation operators over coherent states \cite{agarwal1991nonclassical}, whose expansion leads to the general expression 
\begin{equation}
\label{ph-add-expval-expanded}
\langle \hat x(\alpha) \rangle_n
= \sqrt{2}\,\Re(\alpha)\,
\frac{L_{n}^{(1)}\!\bigl(-|\alpha|^2\bigr)}
     {L_{n}\!\bigl(-|\alpha|^2\bigr)} = \sqrt{2}\,\Re(\alpha)\;
(n+1) \frac{\displaystyle\sum_{k=0}^n
   \frac{|\alpha|^{2k}}{(k+1)\,(n-k)!\,(k!)^2}}
     {\displaystyle\sum_{k=0}^n
   \frac{|\alpha|^{2k}}{(n-k)!\,(k!)^2}}\,.
\end{equation}

Explicitly, the simplest operational nonlinearity is generated by a single-photon addition ($n=1$) which brings the input coherent state $\ket{\alpha}$ to
\begin{equation}
\label{single-ph-add-nonlinearity}
    \langle \hat x(\alpha) \rangle_1 = \sqrt{2} \Re \Big(\underbrace{\alpha}_{\text{Linear map}} + \underbrace{\frac{\alpha}{1+|\alpha|^2}}_{\text{Nonlinear bump}} \Big).
\end{equation}

On the other hand, consider the non-Gaussian state of Eq. \eqref{ph-sub-coherent-wf} but with a photon addition instead of subtraction
\begin{equation}
\label{ph-add-sq-coherent}
\ket{\psi(\alpha,r)} = \hat a^\dagger \hat S(r)\hat D(\alpha)\ket{0} = \hat a^\dagger \ket{r\alpha} \text{ with } \alpha,r \in \mathbb{R}.
\end{equation}
The expectation value of $\hat x$ in such a state has the form
\begin{equation}
\label{nonlin-ph-add-sq-coherent}
    \langle \hat x(\alpha,r) \rangle
    = 
    \sqrt{2} \Re \Big(\underbrace{e^{r}\,\alpha}_{\text{Linear map of }\alpha}
    \;+\;
    \underbrace{
    \frac{\alpha \cdot e^{2r}\cosh r}{\alpha^{2} \cdot e^{2r} + \cosh^{2} r}}_{\text{Rational bump of }\alpha}\Big).
\end{equation}

\begin{figure}[ht]
\centering
\includegraphics[width=0.335\columnwidth, angle=0]{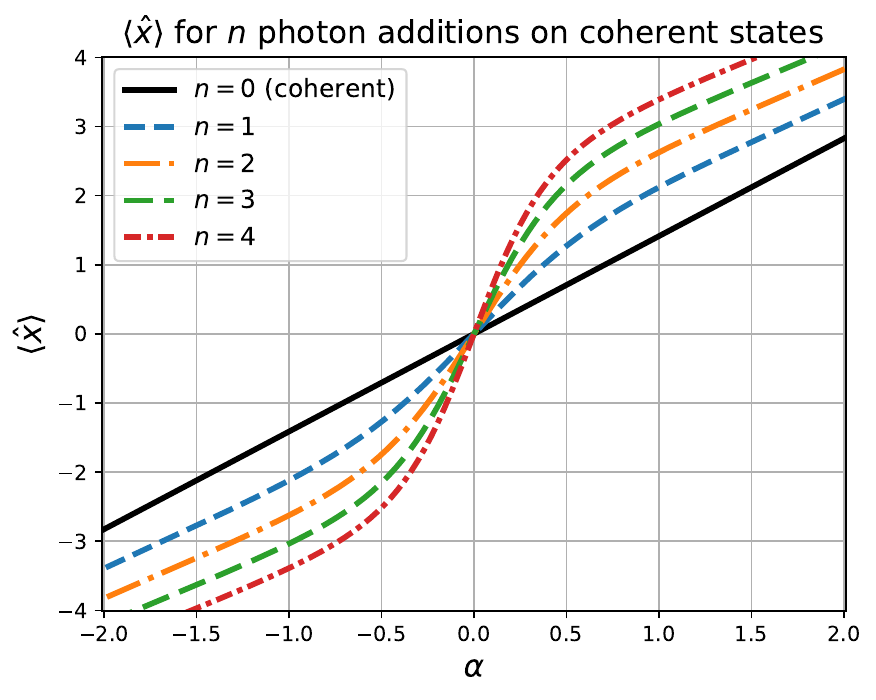}
\includegraphics[width=0.32\columnwidth, angle=0]{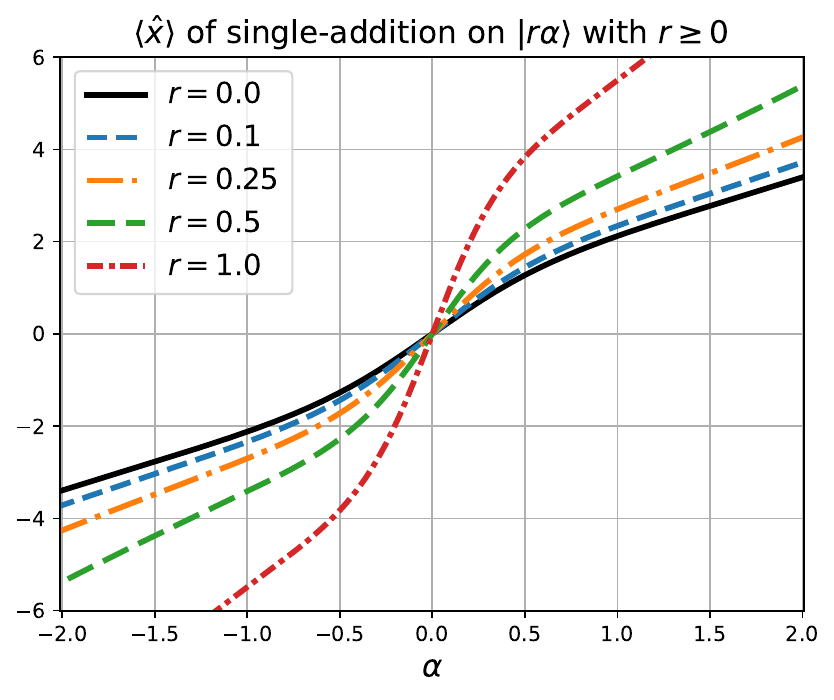}
\includegraphics[width=0.32\columnwidth, angle=0]{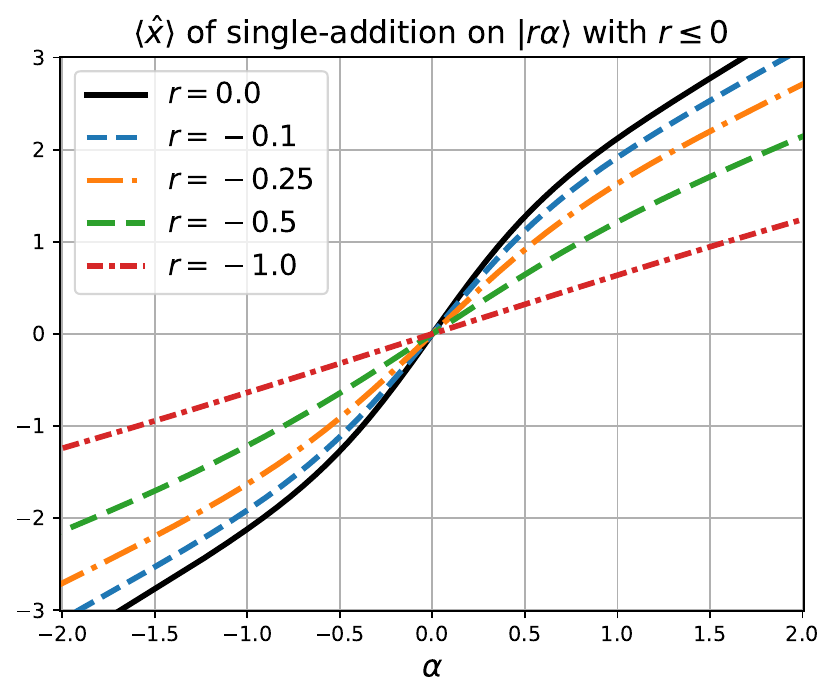}
\caption{Nonlinearity induced by creation operator in $\hat x$ quadrature of squeezed-coherent states with respect to their real displacements $\alpha \in (-2,2)$. Left: $\langle \hat x(\alpha) \rangle_n$ from Eq. \eqref{ph-add-expval-expanded} for $n=0..4$ photon additions. Center and right: $\langle \hat x(\alpha,r) \rangle$ from Eq. \eqref{nonlin-ph-add-sq-coherent} for different values of squeezing in momentum (center, $r \geq 0$) and position (right, $r \leq 0$).}
\label{fig:ph-add-coherent-states}
\end{figure}

From the explicit shapes of Eq. \eqref{ph-add-expval-expanded} shown in the left plot of Fig. \ref{fig:ph-add-coherent-states}, one can clearly notice the smooth nonlinear sigmoid-like effect of the photon addition operations---bounded and non-polynomial---over a real coherent state when the displacement is relatively small, even without squeezing. However, the effect tends to be linear when the displacement is more pronounced, making it a saturated nonlinearity.

The effect of the photon addition is similar when the initial coherent state is squeezed previously to the addition, as seen in Eq. \eqref{nonlin-ph-add-sq-coherent}. The activation function is also non-polynomial and smooth. The center plot of Fig. \ref{fig:ph-add-coherent-states} shows how squeezing momentum quadrature pronounces the nonlinearity induced by the photon creation in the position quadrature mean, while squeezing position (right plot) tends to linearize such effect when the---absolute value---squeezing factor enlarges.

For both cases, the activation functions emerging from the creation operator $\hat a^\dagger$ obey the needs to achieve the UAT, up to some possible limitations.

\section{Exact classical simulation of the CV-QONN model}
\label{appendix:classical-sim}
To classically simulate the QONN from Fig. \ref{fig:physical-qnn} using the Wick--Isserlis expansion of Eq. \eqref{perf_match}, all the Gaussian operators from the layers have to be dragged to the beginning of the circuit for the state to remain Gaussian and, afterwards, apply all the non-Gaussianity which would become a combination of creation and annihilation operators.

This is achieved by the Bogoliubov unitary action of the general Gaussian operator $\hat G$ of the form of Eq. \eqref{general-gauss-lin-quad} over the annihilation operator \cite{ulysse2021classical}
\begin{equation}
\label{unit_act_gaus_lad}
    \hat{\mathbb{S}}_k = \hat{G}\hat{a}_k\hat{G}^\dagger = \bm\delta_k + \sum_{j=1}^N U_{k,j}\hat{a}_j + V_{k,j}\hat{a}_j^\dagger,
\end{equation}
where $N$ is the number of optical modes of the system, $k$ is the mode the operator acts on, $\bm\delta_k \in \mathbb{C}$ is the $k$-th element of the displacement vector associated to $\hat{D}_{\bm \delta}$, and $U_{k,j}$ and $V_{k,j}$ represent entries of the Bogoliubov matrix representing $\hat Q$ as shown in Eq. \eqref{quad-gaussian-op}.

\begin{figure}[ht]
\centering
\includegraphics[width=1\columnwidth, angle=0]{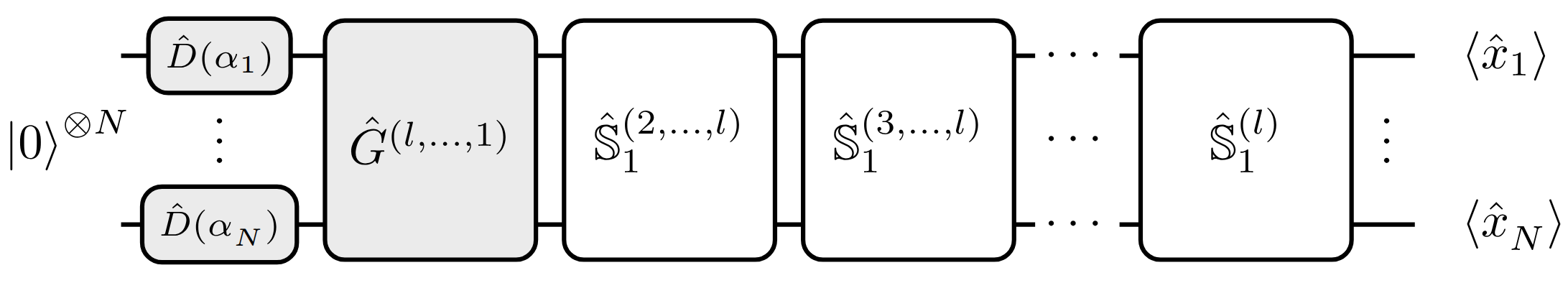}
\caption{Equivalent QONN circuit of Fig. \ref{fig:physical-qnn} adapted for classical simulation, where $\hat{G}^{(l,...,1)} = \hat{G}^{(l)} ... \hat{G}^{(1)}$ is the product of all layers' Gaussians, and $\hat{\mathbb{S}}_1^{(2,...,l)}$ is the $N$-mode ladder superposition operator of Eq. \eqref{unit_act_gaus_lad} whose associated displacement vector $\bm \delta$ and Bogoliubov matrix $B$ correspond to those forming $\hat{G}^{(2,...,l)}$.}
\label{fig:simulation-qnn}
\end{figure}

This way, Wick--Isserlis theorem from Eq. \eqref{perf_match} can be used---along with the observable $\hat x$ over the system modes to be measured---to compute the QONN outcomes.

As an illustrative example for the classical simulation, let's provide a complete construction, description and training of a two-layer QONN with $N=2$ modes, where the dataset has two inputs $\bm \alpha = (\alpha_1, \alpha_2)$ and one output $\langle \hat x_1\rangle$. Its final quantum state is defined by the expression
\begin{equation}
\label{2layer-qnn}
     \hat{a}_1\hat{G}^{(2)} \hat{a}_1\hat{G}^{(1)} \  
    \ket{\bm \alpha}\bra{\bm \alpha}^{\otimes 2}
     \hat{G}^{(1)\dagger} \hat{a}_1^{\dagger} \hat{G}^{(2)\dagger} \hat{a}_1^{\dagger}, 
\end{equation}
which, using Eq. \eqref{unit_act_gaus_lad} for simulation, is equivalent to 
\begin{equation}
\label{2layer-commuted-qnn}
    \hat{a}_1\hat{\mathbb{S}}_1^{(2)} \hat{G}^{(2,1)} \  
    \ket{\bm \alpha}\bra{\bm \alpha}^{\otimes 2}
     \hat{G}^{(2,1)\dagger} \hat{\mathbb{S}}_1^{(2)\dagger} \hat{a}_1^{\dagger} \text{ with } \hat{G}^{(2,1)} = \hat{G}^{(2)}\hat{G}^{(1)}
\end{equation}
Regarding the output acquisition, the observable to be measured is the first-mode position operator $\hat x_1$, whose representation in Fock space is
\begin{equation}
\label{mode1_pos_op}
    \hat x_1 = \frac{\hat a_1 + \hat a^\dagger_1}{\sqrt{2}}.
\end{equation}

First, the initial Gaussian state is the normalized vacuum state of $N=2$ modes defined as in Eq. \eqref{vacuum-st}. Then, uploading the inputs $\bm \alpha$ into the QONN carries the vacuum state to the real coherent state
\begin{equation}
    \ket{\bm \alpha}^{\otimes 2} = \hat{D}_{\bm{\alpha}} \ket{0}^{\otimes 2}: 
    \begin{cases}
        \bm{\bar{b}}_1 = (\alpha_1, \alpha_2, 0, 0) \\
        \Sigma_1 = \Sigma_0 = \frac{1}{4} \mathbb{1}_{4}
    \end{cases}
\end{equation}

After the input encoding stage, the tunable Gaussian operator $\hat{G}^{(1)}$ is constructed composing the quadratic and the linear Gaussian transformations as defined in Eq. \eqref{general-gauss-lin-quad}.

On the one hand, the quadratic Gaussian $\hat{Q}^{(1)}$, of the form of Eq. \eqref{quad-gaussian-op} represented by the Bogoliubov matrix $B^{(1)}$, is built using the Bloch-Messiah decomposition: two different unitary matrices $U_1^{(1)}$ and $U_2^{(1)}$---standing for the passive optics---are assembled using unitary matrices with their corresponding $N^2=4$ free real parameters, which are tuned in the training process, whilst the diagonal matrix $R^{(1)}$ representing the squeezing is built straight-forward by $N=2$ real and positive parameters ($\bm\vartheta = 0$).

On the other hand, the ensemble of displacement operators $\hat{D}_{\bm \delta}^{(1)}$ forming the linear Gaussian is made of $2N=4$ parameters, one for each quadrature of each optical mode.

Then, applying $\hat G$ in Eq. \eqref{2layer-commuted-qnn} brings the quantum system to the Gaussian state 
\begin{equation}
    \rho_G = \hat{G}^{(2,1)} \  
    \ket{\bm \alpha}\bra{\bm \alpha}^{\otimes 2}
     \hat{G}^{(2,1)\dagger} \equiv
     \begin{cases}
        \bm{\bar{b}} = B^{(2,1)} \bm{\bar{b}}_1 + \bm d^{(2,1)} \\
        \Sigma = B^{(2,1)}\Sigma_1 \ B^{(2,1)\dagger},
     \end{cases}
\end{equation}
achieving the linear transformation over $\bm \alpha$ in the means vector, as expected in Eq. \eqref{nn_singlelayer}.

Next, to perform the nonlinear transformation, let's dive into the non-Gaussian phase where the ladder superposition operator $\hat{\mathbb{S}}^{(2)}_1$ is applied to $\rho_G$, moving the quantum system to
\begin{equation}
\label{final-nongauss-state}
    \rho_{NG} = \frac{\hat{a}_1\hat{\mathbb{S}}_1^{(2)} \ \rho_G \ \hat{\mathbb{S}}_1^{(2)\dagger}\hat{a}_1^{\dagger}}{\text{Tr}\big[\hat{a}_1\hat{\mathbb{S}}_1^{(2)} \ \rho_G \ \hat{\mathbb{S}}_1^{(2)\dagger}\hat{a}_1^{\dagger}\big]},
\end{equation}
which is the QONN normalized---non-Gaussian---final state of Eq. \eqref{2layer-commuted-qnn}.

According to Eq. \eqref{unit_act_gaus_lad}, the expansion of $\hat{\mathbb{S}}^{(2)}_1$ in the expression is given by
\begin{equation}
\begin{gathered}
\label{expanded_lad_superpos}
    \hat{a}_1\hat{\mathbb{S}}_1^{(2)} \ \rho_G \ \hat{\mathbb{S}}_1^{(2)\dagger}\hat{a}_1^{\dagger} =\\
    \hat{a}_1(\bm{\delta}_1^{(2)} + B_{1,1}^{(2)}\ \hat{a}_1 + B_{1,2}^{(2)}\ \hat{a}_2
    + B_{1,3}^{(2)}\ \hat{a}_1^\dagger + B_{1,4}^{(2)}\ \hat{a}_2^\dagger)\ \rho_G\ 
    (\bm{\delta}_1^{(2)*} + B_{1,1}^{(2)\dagger}\ \hat{a}_1^\dagger + B_{1,2}^{(2)\dagger}\ \hat{a}_2^\dagger
    + B_{1,3}^{(2)\dagger}\ \hat{a}_1 + B_{1,4}^{(2)\dagger}\ \hat{a}_2)\hat{a}_1^{\dagger}.
\end{gathered}
\end{equation}

Lastly, the QONN output $\tilde f(\bm\alpha)$ is obtained by homodyne measurements of the final state, i.e. adding the observable $\hat x_1$ in the form of Eq. \eqref{mode1_pos_op} to $\rho_{NG}$ and computing its expectation value as
\begin{equation}
\begin{split}
\label{exp_val-single-layer-example}
    \tilde f(\bm\alpha) = \text{Tr} \left[ \hat x_1 \rho_{NG}\right] = \frac{\text{Tr} \left[ \frac{(\hat a_1 + \hat a^\dagger_1)}{\sqrt{2}} \hat{a}_1\hat{\mathbb{S}}_1^{(2)} \ \rho_G \ \hat{\mathbb{S}}_1^{(2)\dagger} \hat{a}_1^{\dagger} \right]}{\text{Tr}\big[\hat{a}_1\hat{\mathbb{S}}_1^{(2)} \ \rho_G \ \hat{\mathbb{S}}_1^{(2)\dagger} \hat{a}_1^{\dagger}\big]}
    =\frac{\sqrt{2}\cdot \Re \Big(\text{Tr} \left[ \hat{\mathbb{S}}_1^{(2)\dagger} \hat{a}_1^{\dagger} \hat{a}_1 \hat a_1 \hat{\mathbb{S}}_1^{(2)} \rho_G \right]\Big)}{\text{Tr}\big[\hat{\mathbb{S}}_1^{(2)\dagger} \hat{a}_1^{\dagger} \hat{a}_1 \hat{\mathbb{S}}_1^{(2)} \rho_G\big]},
\end{split}
\end{equation}
where $\hat{\mathbb{S}}_1^{(2)}, \hat{\mathbb{S}}_1^{(2)\dagger}$ are expanded as Eq. \eqref{expanded_lad_superpos} arising multiple trace subexpressions. The loop perfect matchings equation from Eq. \eqref{perf_match} has to be applied independently on each subexpression, allowing the computation of the full expectation value that represents the QONN output $\tilde f(\bm\alpha)$.

In the classical simulation, this process is repeated during the QONN training choosing an appropriate loss function and optimizing the parameters of $B^{(1)},B^{(2)}$ and $\bm d^{(1)},\bm d^{(2)}$ that build up the quantum optical components in $\hat G^{(1)}$ and $\hat G^{(2)}$.

The computational complexity of the QONN classical simulation is centered in the non-Gaussian stage, which produces a combinatorial explosion due to the Wick--Isserlis expansion.

Since the ladder superposition operator of Eq. \eqref{unit_act_gaus_lad} brings the quantum system to a superposition of different non-Gaussian states---as exemplified in Eq. \eqref{expanded_lad_superpos}---, the expectation value expression for one QONN output Eq. \eqref{exp_val-single-layer-example} will be made of a sum of terms whose trace will be calculated separately and summed at the end.

The number of terms of the general expectation value expression grows with the amount of: optical modes $N$, layers $L$, ladder operators per layer $K$, observables or outputs $O$ as follows
\begin{equation}
\label{num_trace_expressions}
    |\text{Trace expressions}| = O \cdot (2N + 1)^{2K(L-1)}.
\end{equation}

Each of these trace expressions has the form of Eq. \eqref{perf_match} and is computed by Wick--Isserlis using the set $\mathcal{P}$ of all loop perfect matchings of the ladder operators in the expression. For an expression with $M$ ladder operators, the total number of different matchings increases combinatorially guided by Eq. \eqref{num_lpms}.

Note that, additionally, this needs to be applied for the normalization factor of the QONN's final quantum state $\rho_{NG}$ of Eq. \eqref{final-nongauss-state} with $O=1$ in Eq. \eqref{num_trace_expressions}.

\section{Exact expression of a single-layer QONN with two-mode subtraction}
\label{app:two-mode-subtraction}
Let $\rho_G$ be the pre-subtraction Gaussian state of an $N$-mode single-layer QONN represented on the ladder basis
$\bm{\hat b}=(\hat{\bm a},\hat{\bm a}^\dagger)^{\mathsf T}$ by the complex
means $\bm{\bar b}=\langle\bm{\hat b}\rangle$ and the covariance $\Sigma \;=\; \big\langle \delta\bm{\hat b}\,\delta\bm{\hat b}^{\dagger}\big\rangle$ where $\delta\bm{\hat b}=\bm{\hat b}-\langle\bm{\hat b}\rangle$.
The entries of $\Sigma$ are directly used in this ordering
\begin{equation}
  \Sigma_{N+r,N+s}=\langle\delta\hat a_s^\dagger\,\delta\hat a_r\rangle,\qquad
  \Sigma_{r,N+s}=\langle\delta\hat a_r\,\delta\hat a_s\rangle,\qquad
  \Sigma_{r,s}=\langle\delta\hat a_r\,\delta\hat a_s^\dagger\rangle,\qquad \text{with }r,s\in\{1,\dots,N\} .
  \label{eq:index-map}
\end{equation}

In the state $\rho_G$, one independent photon subtraction is performed on each of the optical modes $K=\{p,q\}$ with $p,q\in\{1,\dots,N\}$ through the operator of Eq. \eqref{multi-subtraction-op} built as $\hat A_K=\hat a_p\hat a_q$, yielding the state
\begin{equation}
    \rho=\frac{\hat a_p\hat a_q \rho_G \hat a_q^\dagger \hat a_p^\dagger}{\text{Tr}[\hat a_p^\dagger \hat a_q^\dagger \hat a_p\hat a_q \rho_G]}.
\end{equation}
Then, the readout $\hat x_j=\tfrac{1}{\sqrt2}(\hat a_j+\hat a_j^\dagger)$ over $\rho$ is calculated for an arbitrary mode $j$, not necessarily distinct from $p,q$. This is
\begin{equation}
\label{eq:xj-expanded}
  \langle \hat x_j\rangle_{\rho}
  = \frac{\text{Tr}\big[\hat a_q^\dagger \hat a_p^\dagger \hat a_j \hat a_p \hat a_q \rho_G\big] + \text{Tr}\big[\hat a_q^\dagger \hat a_p^\dagger \hat a_j^\dagger \hat a_p \hat a_q \rho_G\big]}{\sqrt{2}\,\text{Tr}\big[\hat a_q^\dagger \hat a_p^\dagger \hat a_p \hat a_q \rho_G\big]} = \frac{\sqrt{2} \Re\,\text{Tr}\big[ \hat a_q^\dagger \hat a_p^\dagger \hat a_j \hat a_p \hat a_q \rho_G\big]}{\text{Tr}\big[\hat a_q^\dagger \hat a_p^\dagger \hat a_p \hat a_q \rho_G\big]},
\end{equation}
with $\Re$ representing the real part. Now, let $\mathcal D_{\rho}:=\text{Tr}\big[\hat a_q^\dagger \hat a_p^\dagger \hat a_p \hat a_q \rho_G\big]$ denote the expression of the expectation value---trace---in the denominator, which represents the normalization factor of $\rho$ (and the subtraction success probability), and let $\mathcal M_{j|\rho}:=\text{Tr}\big[ \hat a_q^\dagger \hat a_p^\dagger \hat a_j \hat a_p \hat a_q \rho_G\big]$ be the trace expression of the numerator, so
\begin{equation}
\label{eq:xj-final}
  \langle \hat x_j\rangle_{\rho}
  = \frac{\sqrt{2} \Re\,\mathcal{M}_{j|\rho}}{\mathcal{D}_\rho}.
\end{equation}

The heralding denominator $\mathcal D_\rho$ is the fourth-order Gaussian moment which, expanding about the means and applying Wick--Isserlis with the identities of Eq. \eqref{eq:index-map}, gives the closed form
\begin{equation}
  \boxed{\;
  \begin{aligned}
  \mathcal{D}_{\rho}
  &= \big(|\bm{\bar b}_p|^2+\Sigma_{N+p,N+p}\big)\big(|\bm{\bar b}_q|^2+\Sigma_{N+q,N+q}\big)
     + \big|\Sigma_{N+p,N+q}\big|^2 + \big|\Sigma_{p,N+q}\big|^2
  \\[-2pt]&\quad
     + \bm{\bar b}_p\,\bm{\bar b}_q^\ast\,\Sigma_{N+q,N+p}
     + \bm{\bar b}_p^\ast\,\bm{\bar b}_q\,\Sigma_{N+p,N+q}
     + \bm{\bar b}_p\,\bm{\bar b}_q\,\Sigma_{N+q,p}
     + \bm{\bar b}_p^\ast\,\bm{\bar b}_q^\ast\,\Sigma_{p,N+q}\, .
  \end{aligned}\;}
  \label{eq:Dpq-Sigma}
\end{equation}

The numerator $\mathcal M_{j|\rho}$ represents a fifth-order Gaussian moment, where the Wick–Isserlis expansion with nonzero means yields exactly $26$ terms grouped as
\begin{equation}
\label{eq:exp-val-numerator}
\boxed{
    \mathcal M_{j|\rho}
    = \underbrace{\bm{\bar b}_j\,\mathcal D_{\rho}}_{\text{(i) $j$ as a mean: 10 terms}}
    \;+\; \underbrace{\mathcal S^{(2)}_{j|\rho}}_{\text{(ii) 2 fluctuations: 4 terms}}
    \;+\; \underbrace{\mathcal S^{(4)}_{j|\rho}}_{\text{(iii) 4 fluctuations: 12 terms}}.}
\end{equation}

\paragraph*{(i) Mean contribution of mode $j$ (10 terms).} The mean $\bm{\bar b}_j=\langle\hat a_j\rangle$ is coupled with each term in the norm $\mathcal D_\rho$. 

\paragraph*{(ii) Two-fluctuation contribution (4 terms).}
Here $\delta\hat a_j$ is paired with exactly one of
$\{\delta\hat a_p^\dagger,\delta\hat a_q^\dagger,\delta\hat a_p,\delta\hat a_q\}$,
and the remaining three operators are replaced by their means:
\begin{equation}
    \boxed{\
    \begin{aligned}
    \mathcal S^{(2)}_{j|\rho}
    &= \bm{\bar b}_q^\ast\,\bm{\bar b}_p\,\bm{\bar b}_q \Sigma_{j,p}\,
     + \bm{\bar b}_p^\ast\,\bm{\bar b}_p\,\bm{\bar b}_q\Sigma_{j,q}\, + \bm{\bar b}_q^\ast\,\bm{\bar b}_p^\ast\,\bm{\bar b}_q \Sigma_{j,N+p}\,
     + \bm{\bar b}_q^\ast\,\bm{\bar b}_p^\ast\,\bm{\bar b}_p\Sigma_{j,N+q}\, .
    \end{aligned}}
\label{eq:S2}
\end{equation}

\paragraph*{(iii) Four-fluctuation contribution (12 terms).}
Pick one operator among $\{\hat a_q^\dagger,\hat a_p^\dagger,\hat a_p,\hat a_q\}$ to contribute its mean, and pair the remaining four
fluctuations into two-point correlators by Wick--Isserlis. The result is
\begin{equation}
\boxed{\
\begin{aligned}
\mathcal S^{(4)}_{j|\rho}
=\ &\bm{\bar b}_q^\ast\Big[
     \Sigma_{j,p}\,\Sigma_{p,N+q}
   + \Sigma_{j,N+p}\,\Sigma_{N+q,N+p}
   + \Sigma_{j,N+q}\,\Sigma_{N+p,N+p}\Big]\\[2pt]
+\ &\bm{\bar b}_p^\ast\Big[
     \Sigma_{j,q}\,\Sigma_{p,N+q}
   + \Sigma_{j,N+p}\,\Sigma_{N+q,N+q}
   + \Sigma_{j,N+q}\,\Sigma_{N+p,N+q}\Big]\\[2pt]
+\ &\bm{\bar b}_p\Big[
     \Sigma_{j,q}\,\Sigma_{N+q,N+p}
   + \Sigma_{j,p}\,\Sigma_{N+q,N+q}
   + \Sigma_{j,N+q}\,\Sigma_{N+q,p}\Big]\\[2pt]
+\ &\bm{\bar b}_q\Big[
     \Sigma_{j,q}\,\Sigma_{N+p,N+p}
   + \Sigma_{j,p}\,\Sigma_{N+p,N+q}
   + \Sigma_{j,N+p}\,\Sigma_{N+q,p}\Big].
\end{aligned}}
\label{eq:S4}
\end{equation}

Thus, substituting Eqs. \eqref{eq:Dpq-Sigma}, \eqref{eq:S2} and \eqref{eq:S4} into Eq. \eqref{eq:exp-val-numerator}, and inserting the resulting $\mathcal M_{j|\rho}$ together with Eq. \eqref{eq:Dpq-Sigma} into Eq. \eqref{eq:xj-final}, yields the exact normalized output $\langle \hat x_j\rangle_{\rho}$ in closed form, expressed solely in terms of $\rho_G$'s statistics $(\bm{\bar b},\Sigma)$.

If the photon subtractions implemented by $\hat A_K=\hat a_p\hat a_q$ act on two modes that are mutually uncorrelated in $\rho_G$ (i.e., the inter-click correlations vanish, $\Sigma_{N+p,N+q}=\Sigma_{p,N+q}=0$), then the heralding denominator factorizes as $\mathcal D_{\rho}=(|\bm{\bar b}_p|^2+\Sigma_{N+p,N+p})(|\bm{\bar b}_q|^2 + \Sigma_{N+q,N+q})$, and all terms in $\mathcal S^{(4)}_{j|\rho}$ that are proportional to cross-correlations between the clicked modes disappear. In this regime, the normalized readout of mode $j$ contains no inter-click coupling and can be interpreted as the sum of two single-mode (ridge) activations, one associated with mode $p$ and one with mode $q$. In the displacement-free limit for the clicked modes, $\bm{\bar b}_p=\bm{\bar b}_q=0$, one has $\mathcal S^{(2)}_{j|\rho}=\mathcal S^{(4)}_{j|\rho}=0$ and therefore $\langle \hat x_j\rangle_{\rho}=\sqrt{2}\,\Re(\bm{\bar b}_j)$. More generally, if only the anomalous inter-click correlator vanishes ($\Sigma_{p,N+q}=0$ and, thus, $\Sigma_{N+q,p}=0$), then all addends proportional to such terms drop out, and any remaining inter-click dependence is mediated only through the normal correlator $\Sigma_{N+p,N+q}$ (and its Hermitian counterpart $\Sigma_{N+q,N+p}$), together with local variances and correlations involving the readout mode $j$. A single ridge activation is obtained either when one clicked mode is effectively irrelevant for the readout (e.g., $\bm{\bar b}_q=0$ and $\Sigma_{j,q}=\Sigma_{j,N+q}=0$, so only the $p$-activation remains), or in the degenerate case in which both subtraction events act on the same mode ($p=q$, so $\hat A=\hat a_p^2$), which collapses the response to a single (second-order) ridge activation on mode $p$; in this latter case, identical indices introduce combinatorial multiplicities and the corresponding expressions follow by explicitly re-evaluating the Wick pairings for $\hat a_p^2$. Finally, when $j\in\{p,q\}$, additional index coincidences cause some addends to combine accordingly.

\end{document}